\documentclass[prl, aps, nofootinbib, twocolumn, amssymb, superscriptaddress, 10pt]{revtex4-2}
\usepackage{amsmath}
\usepackage{amssymb}
\usepackage{multirow}
\usepackage{amsthm}
\usepackage{amsfonts}
\usepackage{enumerate}
\usepackage{latexsym}
\usepackage{color}
\usepackage{xcolor}
\usepackage{setspace} 
\usepackage{blindtext}
\usepackage{dsfont}
\usepackage{mathrsfs}
\usepackage[normalem]{ulem}

\usepackage{lipsum}

\usepackage{tabularx}
\newcolumntype{Y}{>{\centering\arraybackslash}X}

\usepackage{stackengine}
\usepackage{bbold}

\usepackage{bm}
\usepackage{graphicx}

\usepackage{hyperref}
\hypersetup{
pdfnewwindow=true, colorlinks=true,
linkcolor=blue, anchorcolor=blue,
citecolor=blue, filecolor=blue,
menucolor=blue, urlcolor=blue} 

\usepackage{pifont}

\newcommand{\beginsupplement}{
        \setcounter{table}{0}
        \renewcommand{\thetable}{S\arabic{table}}
        \setcounter{figure}{0}
        \renewcommand{\thefigure}{S\arabic{figure}}
        \setcounter{equation}{0}
        \renewcommand{\theequation}{S\arabic{equation}}
        \setcounter{section}{0}
        \renewcommand{\thesection}{\Alph{section}}
        \setcounter{subsection}{0}
        \renewcommand{\thesubsection}{\arabic{subsection}}
        \setcounter{subsubsection}{0}
        \renewcommand{\thesubsubsection}{\alph{subsubsection}}
}

\newcommand{\vk}{{\mathbf{k}}}

\newcommand{\vq}{{\mathbf{q}}}

\newcommand{\vlr}{{\mathbf{r}}}

\definecolor{amethyst}{rgb}{0.6, 0.4, 0.8}

\begin{document}

\title{Topological chiral superconductivity from antiferromagnetic correlations in
moir\'{e} 
bands with extreme spin-orbit coupling}

\author{Chenyuan Li}
\thanks{\href{cl84@rice.edu}{cl84@rice.edu}}
\affiliation{Department of Physics \& Astronomy,  Extreme Quantum Materials Alliance, Smalley-Curl Institute,
Rice University, Houston, Texas 77005, USA}
\affiliation{Rice Academy of Fellows, Rice University, Houston, Texas 77005, USA}

\author{Fang Xie}
\affiliation{Department of Physics \& Astronomy,  Extreme Quantum Materials Alliance, Smalley-Curl Institute, Rice University, Houston, Texas 77005, USA}
\affiliation{Rice Academy of Fellows, Rice University, Houston, Texas 77005, USA}

\author{Jennifer Cano}
\affiliation{Department of Physics and Astronomy, Stony Brook University, Stony Brook, New York 11794, USA}
\affiliation{Center for Computational Quantum Physics, Flatiron Institute, New York, New York 10010, USA}

\author{Qimiao Si}
\thanks{\href{qmsi@rice.edu}{qmsi@rice.edu}}
\affiliation{Department of Physics \& Astronomy,  Extreme Quantum Materials Alliance, Smalley-Curl Institute, Rice University, Houston, Texas 77005, USA}

\date{\today}

\begin{abstract}
    Motivated by the strong-correlation phenomenology observed near the superconducting phase in twisted bilayer WSe$_2$, we study multi-orbital $t$-$J$ models that are derived from different parameter regimes. The models contain effective antiferromagnetic interactions that  are influenced by the strong underlying spin-orbit coupling.
  The possible superconducting pairing states are investigated in these models.
   We find that the preferred pairing order parameters 
   are associated with the $^{1,2}E$ representations of the three-fold rotation symmetry operator $C_3$, with the $p\pm i p$ component intermixing with the $d\pm id$ component.
    The chiral superconducting states are shown to be topological, based on 
   the Wilson loops of the corresponding Bogoliubov quasiparticles.
    We discuss the implications of our findings for experimental observations,
    as well as the new connections our results uncover between the moir\'{e} superconductivity and its counterpart in bulk quantum materials.
\end{abstract}

\maketitle

{\it \color{blue} Introduction---}
Superconductivity in two-dimensional electronic materials at low carrier densities has emerged as a topic of extensive interest in contemporary condensed matter physics \cite{cao_unconventional_2018, Chen2019Signatures, Park2022Robust, Xia2024Unconventional,Guo2024Superconductivity, Han2025Signatures, Xu2025Signatures}. 
The superconducting states are usually accompanied by correlated insulating phases, or correlated metallic states with spin/valley order.
In particular, the very recent discovery of superconductivity in moir\'{e} transition metal dichalcogenides (TMDs) -- specifically the 
twisted bilayer WSe$_2$ (tWSe$_2$)~\cite{Xia2024Unconventional,Guo2024Superconductivity} -- 
overcame the monopoly of graphene on moir\'{e} superconductivity.
More so than in the moir\'{e} graphene systems, superconductivity here clearly develops in the immediate vicinity of
correlated phases.
The proximity highlights the role of electronic correlations,
as discussed for strongly correlated quantum 
materials~\cite{Kei17.1,Pas21.1,Hu-Natphys2024}.
Accordingly, this discovery of superconductivity in tWSe$_2$, 
along with the associated correlation physics, 
has generated much excitement and considerable theoretical interest
\cite{Xie2024Superconductivity,Kim2025,Christos2024Approximate,Guerci2024Topological,Tuo2024Theory,Fischer2024Theory, Chubukov2025Quantum, Zhu2025Superconductivity, xie2025kondolattice,Schrade2024Nematic,Chen2023Singlet,Zegrodnik2023Mixed,Zegrodnik2024Topological,Zhou2023Chiral,Crepel2024bridging, Qin2024KohnLuttinger}. 

The relevant electronic bands of tWSe$_2$ can be derived from a continuum model, which describes the single-valley electronic structure  \cite{Wu2019topological, Devakul2021Magic}. 
An important recognition is that the underlying bands are expected to be topological
~\cite{Xie2024Superconductivity,xie2025kondolattice,Christos2024Approximate,Guerci2024Topological}, which raises the hope that the ensuing superconductivity is topological.
Weak coupling studies have emphasized the role of a putative critical boson that is Yukawa-coupled to well-defined fermions~\cite{Christos2024Approximate} or the effect of a van Hove singularity located near the Fermi sufface in the bandstructure~\cite{Guerci2024Topological}.
However, the experimental observations at smaller twist angles~\cite{Xia2024Unconventional}, show the high temperature resistivity reaching the Mott-Ioffe-Regel limit (with special bandstructure features such as the van Hove singularity being detached from the Fermi surface). 
As such, the normal state is expected to be strongly correlated. This has led to the notion of topology induced quantum fluctuations~\cite{Xie2024Superconductivity,xie2025kondolattice}. In this description, the natural tendency of a partially flat band developing electronic order can be suppressed by the topology-dictated hybridization of the real-space compact molecular orbitals with the more extended conduction-electron orbitals; accordingly, the system reaches the regime of amplified quantum fluctuations that promote superconductivity. Indeed, it has been shown that this route leads to a superconducting transition temperature that is large as measured by the underlying electronic scale~\cite{Xie2024Superconductivity}, as seen experimentally.

In this work, we investigate the nature of the resulting superconducting state.
We show that, in this strong coupling approach, a topological chiral superconducting state 
robustly develops. The corresponding pairing order parameter is chiral: 
It is associated with the $^{1,2}E$ representations of the three-fold rotation 
symmetry operator $C_3$, with a $p\pm i p$ component intermixing with a $d\pm id$ component.
The chiral edge states are determined and their experimental implications are discussed.
We also show how our work uncovers new connections between the moir\'{e}-TMD superconductivity and its counterpart in bulk systems.

{\it \color{blue} Low-energy physics based on topology-induced quantum fluctuations}
Depending on the choice of the inter-layer and intra-layer moir\'e potentials, 
the top two moir\'e valence bands 
from a single valley can carry either opposite or identical Chern numbers. 
In the case with opposite valley Chern numbers, 
a two-orbital model can be constructed \cite{Xie2024Superconductivity};
in the presence of a vertical displacement field,
one of the orbitals hosts more localized electrons while the other
can be described as carrying itinerant 
conduction electrons. 
In the case with the same Chern number, an alternative approach has
been advanced; it involves a maximally localized orbital,
and another topological power-law orbital which 
carries the entire topological obstruction of both moir\'e bands \cite{xie2025kondolattice}.
In both cases, the more localized orbital with filling factor closer to half filling is expected to be more strongly affected by the Coulomb interaction effects.

Below, we derive two types of Kondo-lattice $t$-$J$ models that are suitable for the above two cases.
Due to the spin-orbit coupling, the spin-spin interactions are no longer SU(2) symmetric, and a Dzyaloshinskii-Moriya term will naturally appear. 
Keeping in mind the regime of amplified quantum fluctuations (as opposed to the regime of flat-band-induced electronic order),
we study the possible pairing channels induced from such spin-exchange interactions.

\begin{figure}[t]
    \centering
    \includegraphics[width=\linewidth]{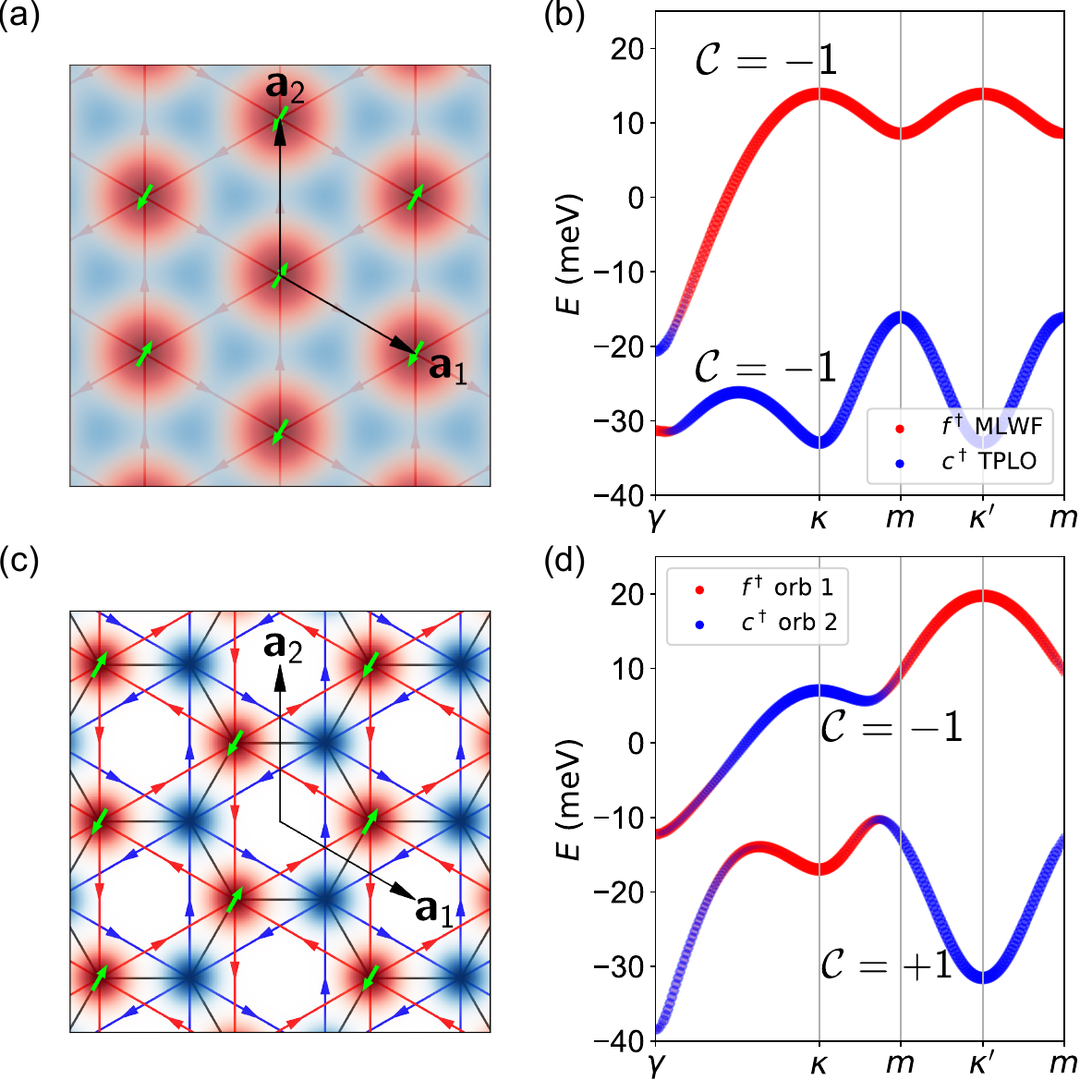}
    \caption{(a) Sketch of the Kondo-lattice $t$-$J$ model of twisted bilayer WSe$_2$ in the $\mathcal{C} = (-1, -1)$ parameter regime.
    The red dots 
    stand for the exponentially localized $f$ orbitals, which hybridize with the topological power-law $c$ orbitals depicted as the bluish background ``clouds''.
    Strong electron interaction leads to spin-valley moments on the $f$ orbitals near half filling, which are represented by the green arrows.
    The spin exchange interactions $J$ are induced by the $f$ orbital hoppings represented by the light red arrows.
    (b) The single-valley bandstructure in the $\mathcal{C}=(-1, 1)$ regime. 
    The color coding (red and blue) stand for the wavefunction weights from the $f$ and $c$ orbitals.
    (c) The sketch of the two-orbital $t$-$J$ model when $\mathcal{C} = (-1, +1)$. 
    In the presence of a non-zero displacement field, orbital $1$ will be much closer to half filling than orbital $2$, leading to localized $f$ spin-valley moments and naturally leading to an effective Kondo-lattice $t$-$J$ model description.
    (d) The single-valley band structure in the $\mathcal{C} = (-1, +1)$ regime.
    The color coding indicates the wavefunction weights of the two exponentially localized orbitals in this parameter regime. 
    }
    \label{fig:sketch}
\end{figure}

{\it \color{blue} Kondo lattice $t$-$J$ models ---} 
When the top two moir\'e bands carry the same valley Chern number, the electronic structure can be described by a maximally localized Wannier function (MLWF) and a topological power-law orbital (TPLO) that carries the full topological obstruction \cite{xie2025kondolattice}.
The real-space structure of such a model is sketched in Fig.~\ref{fig:sketch}(a), and the corresponding band structure is shown in Fig.~\ref{fig:sketch}(b).
with the top band near half-filling, the Coulomb repulsion can lead to spin-valley moments in the MLWF, and the hoppings among them could lead to an effective spin-spin interaction.
Similarly, when the valley Chern numbers are opposite, a two-orbital tight-binding model can also be constructed~\cite{Xie2024Superconductivity}.
In the presence of a perpendicular displacement field, one of the orbitals can also be closer to half-filling than the other, and the Coulomb interaction and the hoppings can also lead to effective spin-valley moments as well as spin-spin interactions, as demonstrated in Fig.~\ref{fig:sketch}(c).
A typical band structure with opposite valley Chern numbers is also shown in Fig.~\ref{fig:sketch}(d).
The details of these models are briefly summarized in the supplemental material (SM) Sec.~A \cite{supplemental_material}.

In both effective models, we denote the orbital with filling factor closer to half-filling as the $f$ orbital, and the other orbital as the $c$ orbital.
The interaction effects are expected to be significantly stronger for the $f$ orbital.
Therefore, an effective Hamiltonian that captures the subspace of the top two bands can be written as:
\begin{align}
    H=H_t+H_{\mathrm{int}}\,,
\end{align}
where the kinetic part is given by:
\begin{align}
    H_{t} & = |t_f| \sum_{\mathbf{R} - \mathbf{R}' = \bm{\delta}_{1,2,3},\sigma} \left( e^{i\sigma \phi_f} f^\dagger_{\mathbf{R}\sigma} f_{\mathbf{R}'\sigma} + {\rm h.c.} \right)
    \nonumber\\
    &+ \sum_{\vk, \sigma} \varepsilon^{(\sigma)}_c(\vk) c^\dagger_{\vk\sigma} c_{\vk\sigma} + \sum_{\vk \sigma} \left(V^{(\sigma)}_{\rm hyb}(\vk) f^\dagger_{\vk\sigma} c_{\vk\sigma} + {\rm h.c.}\right)\nonumber\\
    &-\mu \sum_{\mathbf{R}}(n_{f\mathbf{R}}+n_{c\mathbf{R}})\,,
    \label{eq:kinetic}
\end{align}
in which $\bm{\delta}_{1} = \mathbf{a}_1$, $\bm{\delta}_2 = \mathbf{a}_2$, $\bm{\delta}_3=-\mathbf{a}_{1}-\mathbf{a}_2$ stand for the nearest neighbor lattice vectors of the triangular lattice, as shown in Figs.~\ref{fig:sketch}(a,c).
$|t_f|$ and $\phi_f$ indicate the amplitude and the phase of the nearest neighbor hoppings among the $f$ orbitals.
In particular, in the case with $\mathcal{C} = (-1, -1)$, the value of $\phi_f$ 
can be tuned by the value of the vertical displacement field potential $\varepsilon_D$.
Due to the spin-orbit coupling, such hopping terms can in general be complex numbers with $\phi_f \neq 0 $ or $\pi$.
In both models, the next-nearest neighbor hoppings among $f$ orbitals are smaller than $20\%$ of $|t_f|$, indicating that nearest-neighbor hopping dominates the energetic features of the localized $f$ orbital.
The function $\varepsilon_c(\vk)$ represents the kinetic energy of the corresponding conduction $c$-orbitals in both cases, and $V_{\rm hyb}(\vk)$ stands for the hybridization between the $c$ and $f$ orbitals.
We note that the $c$ orbitals can stand for either the TPLO in the case with the same valley Chern number, or the less localized orbital in the case with opposite valley Chern numbers.

The interaction part, $H_{\mathrm{int}}$, primarily describes the onsite Hubbard interaction $U$
~\cite{Xie2024Superconductivity,xie2025kondolattice}. It has two main effects. First, with $U \gg |t_f|$, $H_{\mathrm{int}}$ leads to exchange interactions between effective $f$- spin-valley moments.
Using the standard Schrieffer-Wolff transformation \cite{Schrieffer1966Relation, Macdonald1988expansion},
the superexchange interaction term between 
the $f$-moments takes the form:
\begin{align}
&H_{\mathrm{int}}=\sum_{\mathbf{R} - \mathbf{R}' = \bm{\delta}_{1,2,3}} J_\perp\left[\left(S^x_{f\mathbf{R}} S^x_{f\mathbf{R}'} + S^y_{f\mathbf{R}} S^y_{f\mathbf{R}'}\right)\right] \nonumber\\
&+J_z\left[ \left(S^z_{f\mathbf{R}} S^z_{f\mathbf{R}'} - \frac14 n_{f\mathbf{R}} n_{f\mathbf{R}'}\right)\right]+D\hat{\bf z}\cdot \left(\mathbf{S}_{f\mathbf{R}} \times \mathbf{S}_{f\mathbf{R}'}\right)\,.
\label{eq:interaction}
\end{align}
Here the $xy$ plane describes where tWSe$_{2}$ lies.
The anisotropic exchange couplings are given by $J_z=J=4|t_f|^2/U$ and $J_\perp=J\cos{2\phi_f}$, and $D=J\sin{2\phi_f}$  represents the Dzyaloshinskii-Moriya term induced by the spin-orbit coupling \cite{Pan2020Band, Chen2023Singlet,Zegrodnik2023Mixed,Zegrodnik2024Topological}. 
The phase factor $\phi_f$ primarily captures the SOC effect, and generically is different from $0$ or $\pi$. In the special case with $\mathcal{C} = (-1, -1)$ and without a displacement field, it
becomes $\pi$ , leading to a vanishing DM term;
still, even in this case, the whole system 
lacks spin SU(2) symmetry, as the hybridization and the kinetic energy of the conduction electrons $c$ are still spin dependent.
Since the next-nearest-neighbor hoppings are significantly smaller than $|t_f|$, 
the corresponding exchange couplings are
negligible.
The detailed derivation of the spin-spin interaction is given in Sec.~B of the SM \cite{supplemental_material}.

The second effect of the Hubbard interaction,
$H_{\mathrm{int}}$, is to renormalize the $t$-hopping of the $f$-electrons. In the following, we will absorb such effects by understanding the corresponding bandwidth $W$ as one that is already renormalized.

\begin{figure*}[t]
    \centering
    \includegraphics[width =\linewidth]{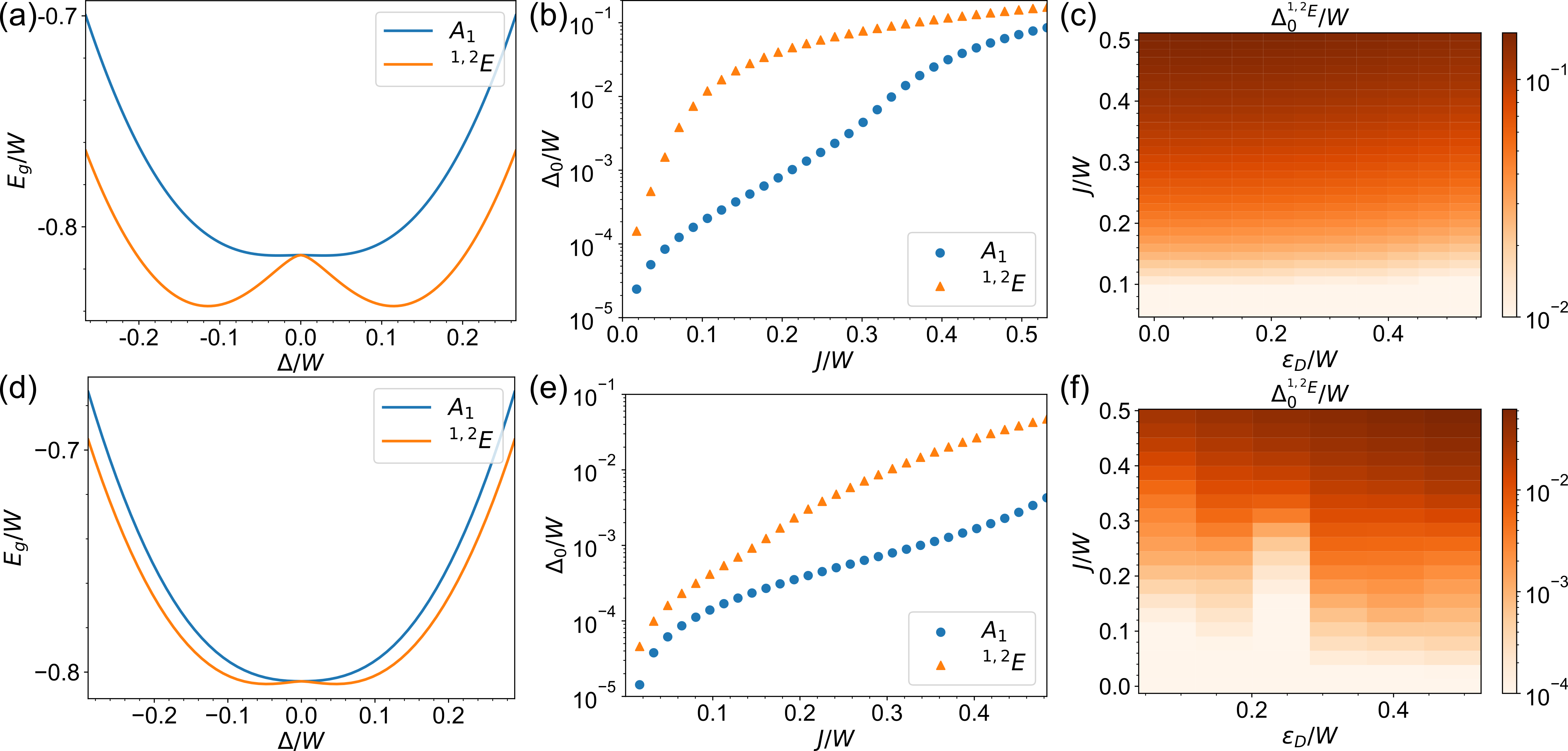}
    \caption{Superconducting properties of the two models with distinct band topology.
    (a-c) correspond to the model with Chern numbers $\mathcal{C}=(-1,-1)$, while (d-f) show results for $\mathcal{C}=(-1,+1)$.
    In both models, all energies are scaled by the bandwidth evaluated at $\varepsilon_D=10$meV: $W=37.6$ meV for $\mathcal{C}=(-1,-1)$ and $W=31.0$ meV for  $\mathcal{C}=(-1,+1)$.
    Unless otherwise specified, the chemical potential $\mu$ is tuned to maintain half-filling. 
    (a,d) Ground state energy density as a function of $\Delta$ for different pairing channels. Parameters used are: (a) $\varepsilon_D/W=0.27$, $J/W=0.40$, and  $\mu/W=0.18$; (d) $\varepsilon_D/W=0.32$, $J/W=0.42$ and $\mu/W=0.18$. In both cases, the results of ${}^{1,2}E$ irreducible representations are degenerate and energetically favored. 
    (b,e) Line cut of the pairing amplitude $\Delta_0$ versus exchange coupling $J$ on a semilog scale at fixed displacement field $\varepsilon_D=10$ meV for two channels. The y-axis shown logarithmically to highlight exponential onset of pairing. 
    (c,f) Superconducting order parameter $\Delta_0$ of the ${}^{1,2}E$ channels for the two models in the $(\varepsilon_D,J)$ plane. The color maps are shown in log scale, and the pairing amplitudes of other channels are set to zero. }
    \label{fig:F23}
\end{figure*}

{\it \color{blue} 
Superconducting pairing---} 
To study pairing instabilities, we consider the effective
interaction derived from Eq.~(\ref{eq:interaction}),
which only includes the opposite-spin pairing term:
\begin{equation}    H_{\mathrm{int}}=\sum_{\mathbf{kk'}}V_\mathbf{kk'}f^\dagger_{\mathbf{k}\uparrow}f^\dagger_{-\mathbf{k}\downarrow}f_{-\mathbf{k'}\downarrow}f_{\mathbf{k'}\uparrow}\,,
\label{eq:pairing term}
\end{equation}
where the pairing potential takes the form $V_{\mathbf{kk'}}=-\frac{2J}{N}\sum_{i=1,2,3}\cos{(k_i+\phi_f)}\cos{(k'_i+\phi_f)}$.
For simplicity we have denoted $k_i\equiv\mathbf{k}\cdot\bm{\delta}_i$ and $N$ is the number of lattice sites. 

In the absence of the vertical displacement field, 
the twisted WSe$_2$ preserves $D_3$ point-group symmetry with a threefold rotation $C_{3z}$ around the out-of-plane axis and 
a twofold rotation $C_{2y}$ that interchanges the two layers. 
When a
vertical displacement field is applied, this $D_3$ symmetry is reduced to $C_3$, which retains only the threefold rotational symmetry. 
Under the $C_3$ symmetry, the superconducting pairing functions can be classified according to 
irreducible representations of the group. In our case, three distinct symmetry channels emerge: the real one-dimensional representation $A_1$,
 and two complex representations ${}^1E$ and ${}^2E$. 
The pairing potential can then be decomposed into these symmetry-adapted basis functions, taking the factorized form 
$V_\mathbf{kk'}=-\frac{2J}{3N}\sum_{\Gamma}\gamma_\Gamma(\mathbf{k})
\bar{\gamma}_\Gamma(\mathbf{k}')$, 
where the sum runs over the irreducible representations $\Gamma=A_1,{}^1E, {}^2E$,
and the pairing channels are described in terms of 
\begin{align}
&\gamma_\Gamma(\mathbf{k})=
\cos{(k_1+\phi_f)} \nonumber\\
&~~~~~~~~~~~+\omega_\Gamma\cos{(k_2+\phi_f)}+\omega_\Gamma^2\cos{(k_3+\phi_f)} \, ,
\\
&{\rm with}~~~\omega_\Gamma=1, \, e^{i2\pi/3}, \, e^{-i2\pi/3}
\, . \nonumber
\label{pairing-channels}
\end{align}
We now perform a mean-field decoupling 
\cite{Goswami2010Superconductivity,Yu2013FeSC,Kotliar88}
of the effective interaction,
Eq.~(\ref{eq:pairing term}), in terms of 
these  pairing channels. The procedure is described in detail 
in Sec.~C of the SM \cite{supplemental_material}.

{\it \color{blue} 
Chiral pairing states--}
We first study the model where the top two bands carry the same non-zero Chern number $\mathcal{C} = (-1, -1)$ \cite{xie2025kondolattice}. 
Our results are shown in Fig.~\ref{fig:F23}(a-c). 
To compare the relative stability of different pairing channels,
we minimize the ground-state energy density $E_g$ for each irreducible representation with respect to its corresponding order parameter, keeping the order parameters of all other channels fixed at zero.
The value of $\Delta$ at which $E_g$ is minimized is taken as the pairing amplitude $\Delta_0$. 
As illustrated in Fig.~\ref{fig:F23}(a), the ${}^{1}E$ and ${}^{2}E$ channels are degenerate and energetically favorable compared to the $A_1$ channel.
To further analyze the evolution of superconductivity, Fig.~\ref{fig:F23}(b) presents a line cut of $\Delta_0$ versus $J$ at fixed displacement field $\varepsilon_D/W=0.27$ on a semilog scale. 
The full phase diagram in Fig.~\ref{fig:F23}(c) displays the dependence of $\Delta_0$ on both $\varepsilon_D$ and $J$ for the ${}^{1,2}E$ channels. 
These results show that superconductivity is more robust
in the ${}^{1,2}E$ channels. 
To verify that the pairing does not mix between degenerate channels, we minimize the ground state energy density (see Eq.~(\ref{eq:free_energy})) as a function of all three order parameters.
We find that the global energy minimum lies along either the ${}^1E$ or ${}^2E$ direction, with no preference for a mixed state, as shown in Sec.~C of the SM \cite{supplemental_material}.

\begin{figure}[t]
    \centering
    \includegraphics[width =\linewidth]{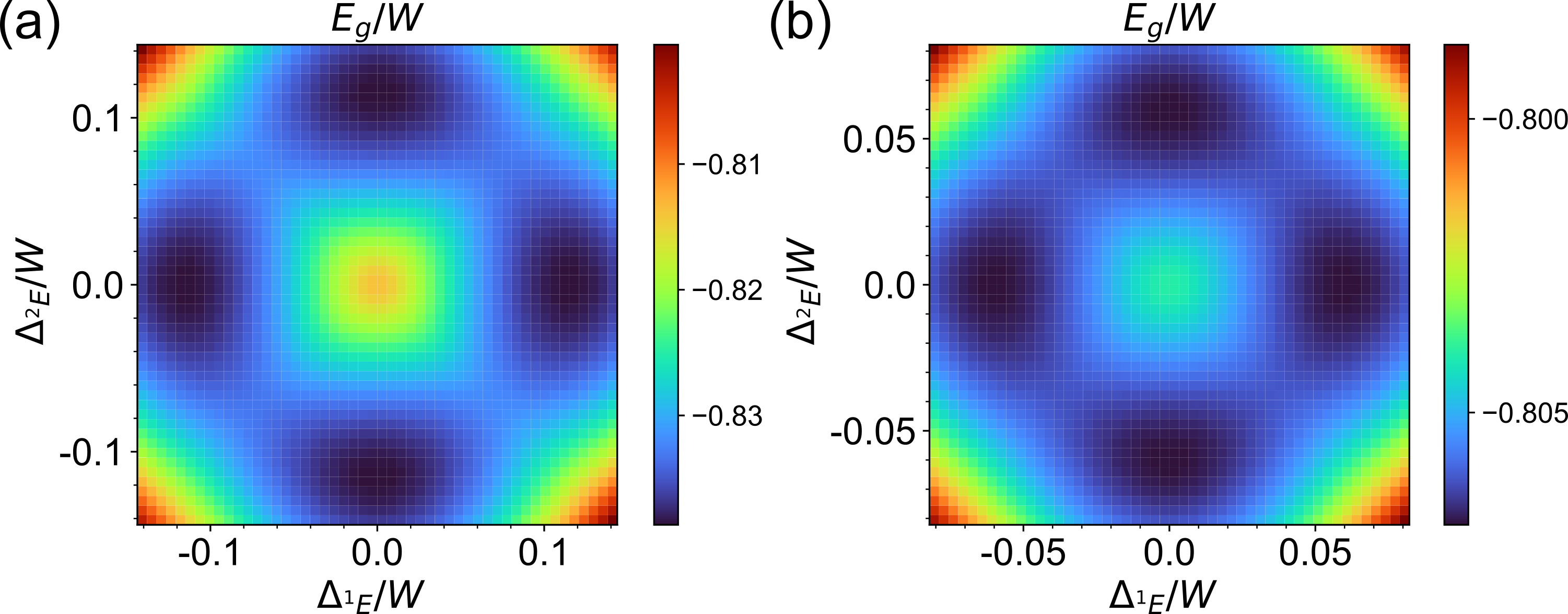}~
    \caption{Ground state energy density $E_g$ as a function of pairing amplitude $\Delta_{{}^1E}$ and $\Delta_{{}^2E}$ for the two models: (a) $\mathcal{C}=(-1,-1)$ (b) $\mathcal{C}=(-1,+1)$. The pairing amplitude of channel $A_1$ is set to be 0, and both $\Delta_{{}^1E}$ and $\Delta_{{}^2E}$ are taken to be real. This figure represents a two-dimensional slice of the full pairing parameter space. }
    \label{fig:Eg_1E2E}
\end{figure}

We now examine the topologically distinct model in which the two orbitals carry opposite Chern numbers, $\mathcal{C}=(-1,+1)$ \cite{Xie2024Superconductivity}. 
Unlike the previous case of $\mathcal{C}=(-1,-1)$, this system requires a finite displacement field $\varepsilon_D$ 
for the Kondo lattice description to apply.
As shown in Fig.~\ref{fig:F23}(d), the ${}^{1,2}E$ pairing channels remain energetically favorable, indicating that the pairing symmetry is robust against the reversal of the orbital Chern number. 
The semilog plot of $\Delta_0$ versus $J$ at fixed $\varepsilon_D$ in Fig.~\ref{fig:F23}(e) again shows that the ${}^{1,2}E$ channels are energetically more favorable than the $A_1$ channel.
As the pairing amplitude is small in the weak-coupling regime, a sufficiently strong exchange interaction is necessary to induce an observable superconducting state.
The phase diagram in Fig.~\ref{fig:F23}(f) is restricted to the finite $\varepsilon_D$ regime, with the modified pairing  and phase boundaries reflect the influence of the underlying band topology. 
The phase $\phi_f$ in this model significantly deviates from $\pi$.
The pairing state contains a large 
$p\pm ip$ component within the ${}^{1,2}E$ representation.

Within the $E$ representation, we further investigate the 
the stability of the chiral ${}^{1}E$ and ${}^{2}E$ superconducting pairng states
against the breaking of further symmetry through a
mixing between the two channels, including any combination that results in
a real order parameter.
Fig.~\ref{fig:Eg_1E2E} shows the ground state energy density Eq.~(\ref{eq:free_energy}) as a function of the pairing amplitudes in ${}^{1}E$ and ${}^{2}E$ channels for both models. 
In each plot, we fix the contribution of channel $A_1$ to 0 and assume that the pairing amplitudes $\Delta_{{}^{1}E}$ and $\Delta_{{}^{2}E}$ are real. 
While the figure represents only a two-dimentional slice of the full parameter space, comprehensive minimization over all pairing amplitudes $\Delta_{A_1}$, $\Delta_{{}^1E}$ and $\Delta_{{}^2E}$ consistently yields the same results: the energy Eq.~(\ref{eq:free_energy}) is minimized in either the pure  ${}^{1}E$ or ${}^{2}E$ channel, without indication of mixing between these two irreps. 
Minimizations with other relative phases between $\Delta_{{}^1E}$ and $\Delta_{{}^2E}$ produce nearly identical energy landscapes, confirming that it is a robust feature.

{\it \color{blue} Topological chiral superconductivity}---
The ${}^{1}E$ and ${}^{2}E$ pairing states represent the intermixing of the $d \pm id$ and $p \pm ip$ channels. Accordingly we explore their topological nature.
We diagnose the BdG state topology using the Wilson loop method, 
the details of which are discussed in Sec.~D of the SM \cite{supplemental_material}.
Since the BdG Hamiltonian can be represented in a form without particle-hole redundancy, and the
${}^{1,2}E$ 
pairing states break time-reversal symmetry, the system falls into class A in the ten-fold classification scheme \cite{Chiu2016classification};
in two dimensions, the topology
can be classified by a $\mathbb{Z}$ index.

For the model in the parameter regime with Chern numbers $\mathcal{C} = (-1, -1)$, the computation of the Wilson loops require a careful consideration of the plane wave decomposition of the TPLO and MLWF.
The calculation shows that the pairing order parameter with irreps $^{1,2}E$ induces Chern number $\pm 2$, which agree with the $d \pm i d$ signature of the corresponding pairing function.
Here the integer  Chern number is obtained by subtracting the Chern number  of the 
BdG superconducting state from that of the underlying normal-state bands.
Generalization to the two-orbital model with $\mathcal{C} = (-1, +1)$ is straightforward. 
Using the order parameters with different $C_3$ irreps, we can also compute the corresponding winding numbers. 
The value of $\phi_f$ for this model is around $2\pi/3$, such that the ${}^{1,2}E$ irreps give rise to $p \pm i p$ pairing functions, as indicated by the $\pm 1$ Wilson loop winding numbers of the corresponding BdG bands.
The detail of the computation of topological indices are discussed in Sec.~D of the SM \cite{supplemental_material}. 

We have studied the edge modes of the Bogoliubov spectrum,
with details given in the SM (Sec.~D)~\cite{supplemental_material}.
There are two chiral modes per edge 
For the case with valley Chern numbers $\mathcal{C} = (-1, +1)$,
there are two chiral modes per edge.
Similarly, we also expect the case with valley Chern numbers $\mathcal{C} = (-1, -1)$ will carry four chiral modes per edge, as suggested by its winding number.
Introducing a Zeeman field may induce
Majorana zero modes
\cite{Sato2010NonAbelian, Black2012Edge}, a detailed 
examination of which is left for future work.

The existence of the chiral edge modes suggests experimental means to probe the topological chiral superconductivity.
In particular, scanning tunneling microscopy (STM) is suited to 
detect such edge modes. More generally, the proposed superconducting state breaks time reversal symmetry, which is 
also amenable to experimental tests through spectroscopic means.

\begin{figure}[t]
    \centering
    \includegraphics[width =\linewidth]{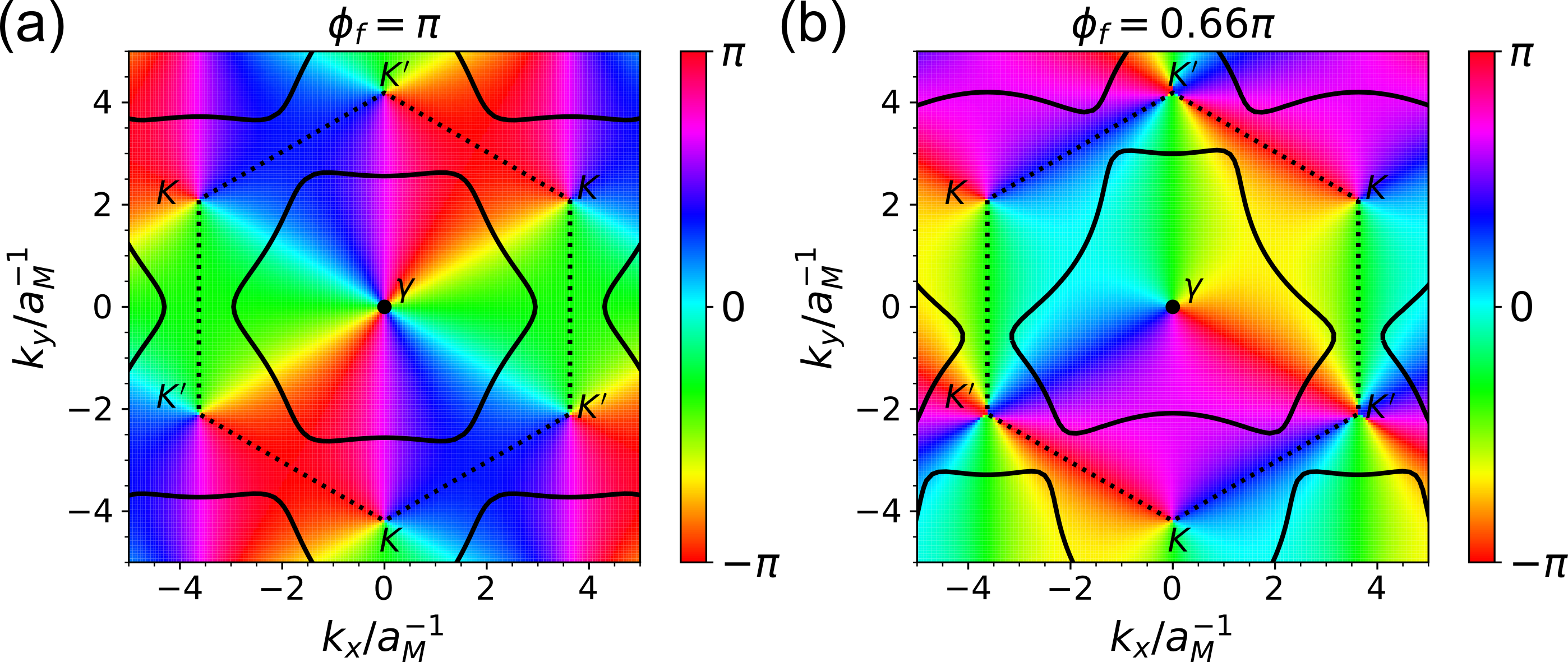}
    \caption{Phase of the pairing function $\gamma_{{}^1E}(\mathbf{k})$.  
    (a) $\phi_f=\pi$ without displacement field in the $\mathcal{C}=(-1,-1)$ model. The phase of $\gamma_{{}^1E}(\mathbf{k})$ winds by $4\pi$ around the $\Gamma$ point, corresponding to a topological winding number of $\pm 2$. 
    (b) $\phi_f$ for $\varepsilon_D=10$ meV in the $\mathcal{C}=(-1,+1)$ model. 
    A $2\pi$ winding of the phase  around the $\Gamma$ point reflects a topological winding number of $\pm 1$. 
    The solid black lines show the Fermi surface and dotted lines mark the first moir\'e Brillouin zone. }
    \label{fig:F4}
\end{figure}

{\it \color{blue} Discussion---}
Several remarks are in order. First, our approach incorporates the strong correlation nature of the normal state~\cite{xie2024kondo, xie2025kondolattice}, as observed experimentally for tWSe$_2$ at relatively small twist angles~\cite{Xia2024Unconventional}. Importantly, the realization of the topological chiral superconductivity is robust provided the underlying bands are topological. By contrast, in weak coupling approaches, the stability of related $E$-type pairing 
requires either the proximity of a van Hove singularity to the Fermi surface~\cite{Guerci2024Topological} or is against a background of 
competitive $A_1$-pairing~\cite{Christos2024Approximate}.

Second, the robust stability of the topological chiral superconductivity we have found in both Kondo lattice $t$-$J$ models raises the prospect that it applies to related moir\'{e} superconductors in which 
the effective degrees of freedom for correlations reside on a triangular lattice. 
While it remains to be proven, the latter proposition
connects well with the stability of the chiral $d+i d$ pairing state in the calculation of the one-band
$t$-$J$ model on a triangular lattice~\cite{Baskaran2003Electronic,Kumar2003Superconductivity,Wang2004Doped,Ogata2003Superconducting,Watanabe2004Variational}, as carried out in the context of the 
experimentally discovered superconductivity in the layered cobaltate 
 ${\rm Na}_{x}{\rm CoO}_{2} \cdot y{\rm H}_{2}{\rm O}$~\cite{Takada-CoO2-SC}.
 As such, our work also reveals new interconnections between the superconductivity of 
 moir\'{e} structures and that of bulk quantum materials. 
 Such a parallel is particularly instructive as it may inspire new experimental means 
 to probe the nature of superconductivity in tWSe$_2$.

{\it \color{blue} Summary---}
To summarize, we have addressed the nature of superconductivity that develops in the twisted bilayer WSe$_2$ at relatively small twist angles. Based on the strong correlation phenomenology observed in these structures and the topological nature of the underlying bands, we have applied the notion of topology-induced quantum fluctuations to derive two Kondo-lattice $t$-$J$ models. The models applied to
different parameter regimes, with  effective interactions that are antiferromagnetic and captures 
the influence of the spin-orbit coupling. We find that the chiral $^{1,2}E$ pairing states are energetically
favored in a robust way. From the Wilson loops of the corresponding Bogoliubov quasiparticles,
we establish the topological nature of the superconducting states. 
Our findings suggest new experimental means to deepen the understanding of these new superconductors, and point to their new connections with the superconductivity observed in the layered cobaltates.
The framework developed here can be extended to other moir\'e materials with nontrivial band topology and strong interactions, raising the prospect to address superconductivity in related systems such as the rhombohedral graphene and twisted MoTe$_2$ \cite{Xu2025Signatures, Han2025Signatures}.
Finally, our work illustrates how lattice symmetry, topology and strong correlations cooperate in providing a robust route towards topological superconductivity.

\begin{acknowledgments}
{\it Acknowledgments.~}
We thank Lei Chen, Yuan Fang, Kin Fai Mak,  Abhay Pasupathy, Jie Shan, and Shouvik Sur for useful discussions. This work has been supported in part by 
the U.S. Department of Energy, Office of Science, Basic Energy Sciences, under Award No. DE-SC0018197 (C.L. and F.X.) and the Robert A.\ Welch Foundation Grant No.\ C-1411 (Q.S.). J.C. acknowledges the support of the National Science Foundation under Grant No. DMR-1942447, support from the Alfred P. Sloan Foundation through a Sloan Research Fellowship and the support of the Flatiron Institute, a division of the Simons Foundation. The majority of the computational calculations have been performed on the Shared University Grid at Rice funded by NSF under Grant No.~EIA-0216467, a partnership between Rice University, Sun Microsystems, and Sigma Solutions, Inc., the Big-Data Private-Cloud Research Cyberinfrastructure MRI-award funded by NSF under Grant No. CNS-1338099, and the Advanced Cyberinfrastructure Coordination Ecosystem: Services \& Support (ACCESS) by NSF under Grant No. DMR170109. Q.S. acknowledges the hospitality of the Aspen Center for Physics, which is supported by NSF grant No. PHY-2210452.
\end{acknowledgments}

\bibliography{reference.bib}
\bibliographystyle{apsrev4-2}

\onecolumngrid
\clearpage

\beginsupplement
\section*{Supplemental Material}
\setcounter{secnumdepth}{3}

\tableofcontents

\section{Continuum model and Wannierization}\label{sec:model}

In this section, we present the continuum model for twisted bilayer transition metal dichalcogenide (TMDC) materials and outline the construction of low-energy effective models via (partial) Wannierization of the top two moir\'e bands in twisted bilayer WSe$_2$.

\subsection{Continuum model}\label{sec:continuum}

The single-valley continuum model \cite{Wu2019topological,Devakul2021Magic} provides a simple description of the low-energy electronic structure of twisted bilayer transition metal dichalcogenides.
For tWSe$_2$, the single-valley electronic structure is captured by the following Hamiltonian:
\begin{equation}
    h_0(\mathbf{r}) = \left(
        \begin{array}{cc}
            \frac{\nabla^2}{2m^*} + \tilde{v}_+(\mathbf{r}) + \frac{\varepsilon_D}{2}  & T(\mathbf{r}) \\
            T^*(\mathbf{r}) & \frac{\nabla^2}{2m^*} + \tilde{v}_-(\mathbf{r}) - \frac{\varepsilon_D}{2}
        \end{array}
    \right)\,,
\end{equation}
in which the two entries of this matrix stands for the electronic states from the top ($\ell = +$) and bottom ($\ell = -$) layers in the same valley.
The effective mass $m^* \approx 0.43 m_e$ describes the low-energy dispersion of the hole pocket near the $K$ and $K'$ points of the single layer Brillouin zone, and $\varepsilon_D$ is the potential energy difference induced by a non-zero vertical displacement field.
$\tilde{v}_\pm(\vlr)$ and $T(\vlr)$ represent the intra- and inter-layer moir\'e potentials:
\begin{align}
    \tilde{v}_\ell(\mathbf{r}) =& 2\tilde{v} \sum_{j=1}^3 \cos\left(\mathbf{g}_j \cdot \mathbf{r} + \ell \psi\right)\,,\\
    T(\mathbf{r}) =& w \sum_{j=1}^3 e^{i\vq_j \cdot \mathbf{r}}\,.
\end{align}
Here the vectors $\mathbf{q}_{1,2,3}$ are the momentum difference between the $K$ points from the top and bottom layers,
and $\mathbf{g}_{1,2,3}$ are three of the shortest moir\'e reciprocal vectors, separated by $120^\circ$ from one another.

\begin{figure}[t]
    \centering
    \includegraphics[width=0.6\linewidth]{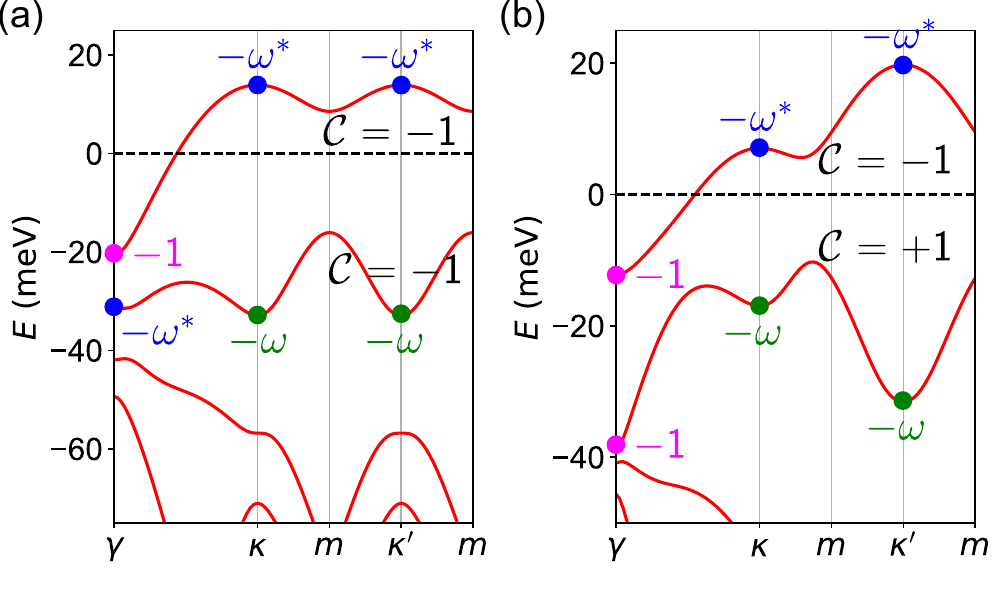}
    \caption{(a) The single-valley band structure of the continuum model at twisting angle $\theta = 3.65^\circ$ with moir\'e potential strengths $\tilde{v} = 9\,\rm meV$ and $w = 18\,\rm meV$. 
    The top two bands carry the same valley Chern number.
    (b) The single-valley band structure of the continuum model with moir\'e potential strengths $\tilde{v} = 13\,\rm meV$, $w = 15\,\rm meV$ and displacement field potential $\varepsilon_D = 15\,\rm meV$. 
    The top two moir\'e bands carry opposite valley Chern numbers.
    The $C_{3z}$ eigenvalues at high symmetry points $\gamma$, $\kappa$ and $\kappa'$ are labeled.
    Here $\omega = e^{i\frac{2\pi}{3}}$.
    }
    \label{fig:tmd_cm}
\end{figure}

The continuum Hamiltonian can be numerical solved by the plane wave basis.
Using these plane wave basis, under which the Bloch wave function of the $n$-th band at $\vk$ will have the following form:
\begin{equation}
    \psi_{n\vk}(\mathbf{r},\ell) = \sum_{\mathbf{Q} \in \mathcal{Q}_\ell} u_{\mathbf{Q},n}(\vk) e^{i(\vk - \mathbf{Q})}\,,
\end{equation}
in which $\mathcal{Q}_\ell$ stands for the momentum space lattice sites from the layer $\ell$ \cite{xie2024kondo, xie2025kondolattice}.
Since the continuum model also possesses a $C_{3z}$ rotation symmetry, we can also compute the $C_{3z}$ eigenvalues of the Bloch state at a given high symmetry point $\mathbf{K}$ as:
\begin{equation}
    \xi_\mathbf{K}(C_{3z}) = e^{i\frac{\pi}{3}}\sum_{\mathbf{Q}} u^*_{C_{3z}\mathbf{Q},n}(C_{3z}\mathbf{K})u_{\mathbf{Q}, n}(\mathbf{K})\,,
\end{equation}
in which $e^{i\frac{\pi}{3}}$ comes from the rotation eigenvalue of spin-$\frac12$ particles.
The $C_{3z}$ eigenvalues of the top two moir\'e bands at these high symmetry points are also labeled in Fig.~\ref{fig:tmd_cm}(d).
The Chern numbers of these two bands at twisting angle $\theta = 3.65^\circ$ are also both $\mathcal{C} = -1$, which agree with the product of the three rotation eigenvalues \cite{Fu2007Topological, Hughes2011Inversion}.

The potential strengths are fitted from {\it ab initio} simulations, and the topological properties of the top moir\'e bands are sensitive to the actual values of these values. 
For example, these parameters are chosen as $\tilde{v} = 9\rm \, meV$, $\psi = 128^\circ$ and $w = 18 \,\rm meV$ in Ref.~\cite{Devakul2021Magic}, and at twisting angle $\theta = 3.65^\circ$, the top two moir\'e bands will have the same Chern number.
With the values of $\tilde{v}$, $w$ changed to $13\,\rm meV$, $15\,\rm meV$, and with a displacement field potential $\varepsilon_D = 15\,\rm meV$, the top two bands will have different valley Chern numbers.
The band structure and high symmetry point rotation eigenvalues of these two cases are also shown in Fig.~\ref{fig:tmd_cm}.

The two effective models described in the main text are based on the previously mentioned two different parameter choices.
In the following, we will discuss these two models with more detail in Secs.~\ref{sec:partial-wannier} and \ref{sec:two-orbital-model}.

\subsection{Partial Wannierization}\label{sec:partial-wannier}

We first investigate the band topology in the case shown in Fig.~\ref{fig:tmd_cm}(a), in which the top two moir\'e bands have the same valley Chern number $\mathcal{C} = (-1, -1)$.
The $C_{3z}$ eigenvalues of the top two moir\'e bands {\it cannot} be written as the summation of two elementary band representations.
Therefore, they cannot be maximally Wannierized together. 
However, one can still notice that if we combine the Bloch state at $\gamma$ point from the second moir\'e band, and the Bloch states away from the $\gamma$ point from the first moir\'e band, then an elementary band representation $(\overline{^2E})_{1a}\uparrow P3$ can be found.

Based on this observation, the authors proposed a ``partial Wannierization'' procedure in Ref.~\cite{xie2025kondolattice}.
Using the ``disentangle'' Wannierization functionality provided by \textsc{Wannier90} \cite{Marzari1997Maximally,Souza2001Maximally,Pizzi2020wannier}, we are able to construct the wave function which indeed corresponds to a maximally localized orbital dubbed MLWF.
Such MLWF is constructed by choosing a proper $\vk$-dependent superposition of the wave functions of the top two moir\'e bands:
\begin{equation}
    f^\dagger_{\vk,\sigma} = \sum_{n=1,2} c^\dagger_{\vk, n, \sigma} U^{\rm dis}_{n}(\vk) = \sum_{\mathbf{Q}} \tilde{u}^{(\sigma)}_{\mathbf{Q},f}(\vk) c^\dagger_{\vk- \mathbf{Q}, \sigma}\,,
\end{equation}
in which $c^\dagger_{\vk, n, \sigma}$ stands for the creation operator of the top two moir\'e energy bands, and $c^\dagger_{\vk-\mathbf{Q}, \sigma}$ stands for the plane wave creation operator for the continuum model, which will also be used in Sec.~\ref{sec:topo-idx}.
Accordingly, the corresponding ``conduction orbital'' can be constructed by a Gram–Schmidt process, which will carry all the topology obstruction with total valley Chern number $\mathcal{C} = -2$.
Hence, it cannot be maximally Wannierized \cite{thouless_wannier_1984}.
We further fix the gauge of this orbital using the method introduced in Ref.~\cite{Xie2024Chern}, which leads to a Wannier function that decays in a polynomial manner.

We can still define the ``kinetic energy'' of the $c$ orbitals, and the hybridization functions between the $f$ and $c$ orbitals.
The effective Hamiltonian which captures the electronic structure of the top two moir\'e bands can be written as the following form:
\begin{equation}
    H_t = \sum_{\mathbf{R}\mathbf{R}'\sigma}t_{ff}^{(\sigma)}(\mathbf{R}) f^\dagger_{\mathbf{R}+\mathbf{R}', \sigma} f_{\mathbf{R}',\sigma} + \sum_{\vk,\sigma} \varepsilon^{(\sigma)}_c(\vk) c^\dagger_{\vk,\sigma} c_{\vk\, \sigma} + \sum_{\vk,\sigma}\left( V_{\rm hyb}^{(\sigma)}(\vk) f^\dagger_{\vk,\sigma} c_{\vk,\sigma} + {\rm h.c.}\right)\,.
\end{equation}
Here, we also note that the nearest neighbor hoppings among $f$ orbital is more than $5$ times larger than the next-nearest neighbor ones.
Besides, since the $c$ orbitals in this model is not maximally localized, we emphasize that this Hamiltonian is {\it not} a tight-binding model in the traditional sense.

\subsection{Two-orbital effective model}\label{sec:two-orbital-model}

We also briefly describe the effective two-orbital model, that is suitable for the parameter regime shown in Fig.~\ref{fig:tmd_cm}(b), in which the top two moir\'e bands carry the opposite valley Chern numbers $\mathcal{C} = (-1, +1)$.
The construction of the two orbital model was addressed in the supplemental material of Ref.~\cite{Xie2024Superconductivity}.
Noticing that in Fig.~\ref{fig:tmd_cm}(b), the $C_{3z}$ eigenvalues of the top two bands at high symmetry points can form two elementary band representations $(\overline{E})_{1b} \uparrow P3$ and $(\overline{E})_{1c} \uparrow P3$ (or equivalently, elementary band representations $(^2 E)_{1b} \uparrow P3$ and $(^2E)_{1c}\uparrow P3$ if we consider the continuum model as a ``spinless'' Hamiltonian).
Therefore, a tight-binding model can be constructed for the top two bands using \textsc{Wannier90} \cite{Marzari1997Maximally,Souza2001Maximally,Pizzi2020wannier} in a straightforward manner.
The effective Hamiltonian can have the following form:
\begin{equation}
    H_t = \sum_{\mathbf{R}\mathbf{R}'\sigma}\left(t_{ff}^{(\sigma)}(\mathbf{R}) f^\dagger_{\mathbf{R}+\mathbf{R}', \sigma} f_{\mathbf{R}',\sigma} + t_{cc}^{(\sigma)}(\mathbf{R}) c^\dagger_{\mathbf{R}+\mathbf{R}', \sigma} c_{\mathbf{R}',\sigma} + t_{fc}^{(\sigma)} f^\dagger_{\mathbf{R}+\mathbf{R}',\sigma}c_{\mathbf{R}',\sigma} + {\rm h.c.}\right)\,.
\end{equation}
The values of all these hopping terms are also provided in Ref.~\cite{Xie2024Superconductivity}. 
For example, the nearest-neighbor hopping among the $f$ orbitals is given by $t^{(\uparrow)}_{ff}(\mathbf{a}_1) = 4.28 e^{i 118^\circ} \, \rm meV$, and the nearest-neighbor hybridization between the two types of orbitals are given by $t^{(\uparrow)}_{fc}(\mathbf{0}) = 3.51 e^{i 48^\circ}\, \rm meV$.
The spin down components ($\sigma = \downarrow$) are simply the complex conjugation of their spin up counterparts due to the time-reversal symmetry.

\subsection{The limit of vanishing displacement field}

The previous discussion about these two effective models are based on space group $P3$, which possesses three-fold rotation symmetry.
In fact, at zero displacement field limit, the moir\'e continuum model also has an extra $C_{2y}T$ symmetry. 
Besides, due to the over-simplification in the choice of the moir\'e potential, the moir\'e continuum model also has a ``pseudo-inversion'' symmetry \cite{Yu2024Fractional}.

The extra symmetries in vanishing displacement field limit imposes extra constraints to the effective model.
For example, the nearest neighbor hopping of MLWF in the case with $\mathcal{C} = (-1, -1)$ is real $(t_f \in \mathbb{R})$ in the limit $\varepsilon_D =0$.
Nonetheless, the effective Kondo lattice and $t$-$J$ model mapping are still valid in this limit, as the MLWF and TPLO are still not equivalent.
In contrast, in the parameter regime with Chern numbers $\mathcal{C} = (-1, +1)$, the two orbitals will become ``degenerate''.
Near the total filling factor $\nu_h = 1$, both orbitals will be around quarter filling instead of half filling.
As a consequence, the Kondo lattice description is not suitable.
Due to this reason, we limit our discussion for the two-orbital model in the presence of displacement field [see Fig.~\ref{fig:Eg_1E2E}(f) in the main text]. 

\section{Derivation of the Kondo lattice \texorpdfstring{$t$-$J$}{t-J} model}\label{sec:tj}
\subsection{Schrieffer-Wolff transformation}

In this section, we briefly review the derivation of the effective $t$-$J$ model from a tight-binding model. The derivation is similar to that in Ref.~\cite{Macdonald1988expansion}, while here we also take the spin-orbit coupling into account as in Refs.~\cite{Pan2020Band,Chen2023Singlet}.
The foundation of the derivation is the Schrieffer-Wolff transformation \cite{Schrieffer1966Relation}, which can be considered as a 2nd order perturbative expansion of a Hamiltonian in a small parameter.
We start with a Hamiltonian of the form:
\begin{equation}
    H = H_0 + H_1\,,
\end{equation}
in which $H_0$ is the unperturbed Hamiltonian with a low-energy subspace $\mathcal{H}_L$ and a high-energy subspace $\mathcal{H}_H$, and $H_1$ is the perturbation.
It is usually more convenient to rewrite the Hamiltonian in the block form:
\begin{equation}
    H_0 = \left(
        \begin{array}{cc}
            H_L & 0 \\
            0 & H_H
        \end{array}
    \right)\,, ~~~ H_1 = \left(
        \begin{array}{cc}
            0 & V \\
            V^\dagger & 0
        \end{array}
    \right)\,,
\end{equation}
in which the matrices $H_L$, $H_H$ are already diagonalized, and the eigenvalues of $H_L$ are well-separated from those of $H_H$.
For simplicity, we assume that the low-energy subspace is spanned by eigenstates with the same eigenvalue $\varepsilon^L$, and the high-energy subspace is spanned by eigenstates with all eigenvalues satisfying $\varepsilon^H_m \geq \varepsilon^L$.
The Sch\"odinger equation of this Hamiltonian is given by:
\begin{align}
    H_L |\psi_L\rangle + V |\psi_H\rangle &= E |\psi_L\rangle \,, \\
    V^\dagger |\psi_L\rangle + H_H |\psi_H\rangle &= E |\psi_H\rangle \,.
\end{align}
We first solve the second equation for $|\psi_H\rangle$:
\begin{equation}
    |\psi_H\rangle = \sum_{m \in \mathcal{H}_H}|m\rangle\frac{\langle m| V^\dagger |\psi_L\rangle}{E - \varepsilon^H_m}\,,
\end{equation}
in which $|m\rangle$ stands for the eigenstates of $H_0$ in the high-energy subspace. 
Substituting this into the first equation, we have:
\begin{equation}
    \left[H_L + \sum_{m \in \mathcal{H}_H} \frac{V|m\rangle \langle m| V^\dagger}{E - \varepsilon_m^H} \right] |\psi_L\rangle = E |\psi_L\rangle\,.
\end{equation}
Obviously, this equation is an effective Sch\"odinger equation for the low-energy subspace with second order perturbation corrections from $H_1$.
Since the low-energy subspace is spanned by eigenstates with the same eigenvalue $\varepsilon^L$, we can approximate the eigenvalue $E$ in the low-energy subspace as $E \approx \varepsilon^L$.
Therefore, in the low-energy subspace $n,n'\in \mathcal{H}_L$, the effective Hamiltonian is given by:
\begin{equation}\label{eqn:Schrieffer-Wolff}
    [H_{\text{eff}}]_{nn'} = \varepsilon^L \delta_{nn'} + \sum_{m\in\mathcal{H}_H} \frac{\langle n | V |m \rangle \langle m | V^\dagger | n' \rangle}{\varepsilon^L - \varepsilon^H_m}\,.
\end{equation}
Here the second term contains the second order perturbation corrections from $H_1$ for the low-energy effective theory.
In the following, we will apply this method to derive the effective $t$-$J$ model from a tight-binding model with strong spin-orbit coupling.

\subsection{Superexchange interaction with spin-orbit coupling}

Generally, the $t$-$J$ model is considered as a low-energy effective theory of a tight-binding model with strong Hubbard interaction.
The spin-spin interactions can be derived from a Schrieffer-Wolff transformation of the above Hamiltonian at half-filling.
Clearly, the low-energy subspace is spanned by the states with exactly one electron on each site, which all have eigenvalues $\varepsilon^L = 0$.
The wave functions of these low-energy states can be written as:
\begin{equation}
    \mathcal{H}_L = \left\{\{|\sigma_{\mathbf{R}\alpha}\}\rangle = \prod_{\mathbf{R}\alpha} c^\dagger_{\mathbf{R}\alpha,\sigma_{\mathbf{R}\alpha}} |0\rangle \Bigg{|} \sigma_{\mathbf{R}\alpha} = \uparrow,\downarrow \right\}\,,
\end{equation}
where $|0\rangle$ is the vacuum state.
Such low-energy subspace is equivalent to the Hilbert space of a spin-1/2 system. 
The coupling between the spin moments on any two different sites can be derived from Schrieffer-Wolff transformation using the hopping terms as the perturbation.
For simplicity, we first solve the two-site problem considering only $\mathbf{R}\alpha$ (denoted as 1) and $\mathbf{R}'\beta$ (denoted as 2).
In the low-energy subspace of two-site Hubbard model, there are four states:
\begin{align}
    |\uparrow \uparrow \rangle =& c^\dagger_{1\uparrow} c^\dagger_{2\uparrow} |0\rangle\,, \\
    |\uparrow \downarrow \rangle =& c^\dagger_{1\uparrow} c^\dagger_{2\downarrow} |0\rangle\,, \\
    |\downarrow \uparrow \rangle =& c^\dagger_{1\downarrow} c^\dagger_{2\uparrow} |0\rangle\,, \\
    |\downarrow \downarrow \rangle =& c^\dagger_{1\downarrow} c^\dagger_{2\downarrow} |0\rangle\,.
\end{align}
Besides, there are two high-energy states, which will be denoted as $|20\rangle$ and $|02\rangle$:
\begin{align}
    |20\rangle =& c^\dagger_{1\uparrow} c^\dagger_{1\downarrow} |0\rangle\,, \\
    |02\rangle =& c^\dagger_{2\uparrow} c^\dagger_{2\downarrow} |0\rangle\,.    
\end{align}
Suppose the hopping from site $2$ to site $1$ with spin $\uparrow$ is given by $t \in \mathbb{C}$, the perturbation Hamiltonian can be written as:
\begin{align}
    H_0 &= U \left(|20\rangle \langle 20 | + |02\rangle \langle 02|\right)\,, \\
    H_1 &= t c^\dagger_{1\uparrow} c_{2\uparrow} + t^* c^\dagger_{2\uparrow} c_{1\uparrow} + t c^\dagger_{2\downarrow} c_{1\downarrow} + t^* c^\dagger_{1\downarrow} c_{2\downarrow}\,.
\end{align}
Using the expression in Eq.~(\ref{eqn:Schrieffer-Wolff}), it is easy to show that only the following matrix elements in the effective Hamiltonian are non-zero:
\begin{align}
    \langle \uparrow \downarrow | H_{\rm eff} |\uparrow \downarrow \rangle &= - \frac{|\langle \uparrow\downarrow|H_t |02 \rangle |^2}{U} - \frac{|\langle \uparrow\downarrow|H_t |20 \rangle |^2}{U} = -\frac{2|t|^2}{U}\,,\\
    \langle \downarrow \uparrow | H_{\rm eff} |\downarrow \uparrow \rangle &= - \frac{|\langle \downarrow\uparrow|H_t |02 \rangle |^2}{U} - \frac{|\langle \downarrow\uparrow|H_t |20 \rangle |^2}{U} = -\frac{2|t|^2}{U}\,,\\
    \langle \uparrow \downarrow | H_{\rm eff} | \downarrow \uparrow \rangle &= -\frac{\langle \uparrow \downarrow | H_t |02\rangle \langle 02 | H_t | \downarrow \uparrow \rangle}{U} - \frac{\langle \uparrow \downarrow | H_t |20\rangle \langle 20 | H_t | \downarrow \uparrow \rangle}{U} = \frac{2t^2}{U}\,,\\
    \langle \downarrow \uparrow | H_{\rm eff} | \uparrow \downarrow \rangle &= \langle \uparrow\downarrow |H_{\rm eff}| \downarrow \uparrow\rangle^* = \frac{2(t^*)^2}{U}\,.
\end{align}
These matrix elements can be reorganized into the following form using the spin operators:
\begin{align}
    H_{\rm eff} =& \frac{4|t|^2}{U} \left[\frac{1}{2}\left(e^{i2\phi} S^+_{1} S^-_{2} + e^{-i2\phi} S^-_{1} S^+_{2} \right) + \left(S^z_{1} S^z_{2} - \frac14\right)\right] \\
    =& \frac{4|t|^2}{U}\left[\left(S^x_1 S^x_2 + S^y_1 S^y_2\right)\cos 2\phi + \hat{\bf z}\cdot \left(\mathbf{S}_1 \times \mathbf{S}_2\right)\sin 2\phi + \left(S^z_1 S^z_2-\frac14\right)\right]\,,
\end{align}
where the phase angle in this effective spin Hamiltonian is defined as $\phi = {\rm arg}\,t$.
Note that the effective spin-spin interaction can also contain a Dzyaloshinskii-Moriya (DM) term when $t^2$ is not a real number, explicitly breaking the $SU(2)$ symmetry.

One may also notice that the effective Hamiltonian of the two-site problem still has the energy spectrum of an $SU(2)$ symmetric Heisenberg model.
This is due to a ``hidden'' $SU(2)$ algebra that only exists in this two-site variant of the model.
We can consider  the following gauge transformation for the fermionic states on the second site:
\begin{align}
    \tilde{c}^\dagger_{2\uparrow} &= c^\dagger_{2\uparrow} e^{-i\phi}\,, \\
    \tilde{c}^\dagger_{2\downarrow} &= c^\dagger_{2\downarrow} e^{i\phi}\,.
\end{align}
Obviously, such transformation will remove the phase factor in the hopping term, and will also lead to the following transformation of the spin operator on site 2:
\begin{equation}
    \tilde{S}^+_{2} = S^+_{2} e^{-i2\phi}\,.
\end{equation}
This is equivalent to a $2\phi$ rotation of the spin on site 2 around the $z$-axis, and the DM interaction term will be eliminated by this transformation.
Therefore, this two-site Hamiltonian will commute with the operators $\tilde{S}^{\pm} = S^\pm_{1} + \tilde{S}^\pm_{2}$ and $\tilde{S}^z = S_1^z + S_2^z$, which form an $SU(2)$ algebra.
However, such transformation cannot be generalized to the lattice case when the hoppings of the lattice model form closed loops.
The only exception is the case when the accumulated rotation angles of the $S^+$ operators along any closed loops are integer multiples of $2\pi$, which corresponds to a tight-binding model with the accumulated phase factors along any closed loops being integer multiples of $\pi$ for a given spin species. 
Such tight-binding model can be transformed into the form with all real hopping amplitudes by a gauge transformation, which will lead to an $SU(2)$ symmetric Hamiltonian.
In generic cases, the effective Hamiltonian will contain DM terms that cannot be removed by any gauge transformation, and  will only have a $U(1)$ symmetry associated with the total $S^z$ operator.

\subsection{\texorpdfstring{$t$-$J$}{t-J} model description for tWSe\texorpdfstring{$_2$}{2}}

As we have already addressed in Sec.~\ref{sec:model}, the top two bands of the continuum model can be well described by compact molecular orbitals, with one orbital near half-filling being more correlated. 
We denote this orbital as the $f$ orbital, while the more itinerant one is labeled as the $c$ band. 
Based on this construction, the effective Hamiltonian for nearest neighbor interactions takes the form as: 
\begin{align}
    H_{t} =& t_f \sum_{\mathbf{R} - \mathbf{R}' = \bm{\delta}_{1,2,3},\sigma} \left( e^{i\sigma \phi_f} f^\dagger_{\mathbf{R}\sigma} f_{\mathbf{R}'\sigma} + {\rm h.c.} \right) + \sum_{\vk, \sigma} \varepsilon^{(\sigma)}_c(\vk) c^\dagger_{\vk\sigma} c_{\vk\sigma} + \sum_{\vk \sigma} \left(V^{(\sigma)}_{\rm hyb}(\vk) f^\dagger_{\vk\sigma} c_{\vk\sigma} + {\rm h.c.}\right)\,,\\
    H_{J} =& \frac{4|t_f|^2}{U} \sum_{\mathbf{R} - \mathbf{R}' = \bm{\delta}_{1,2,3}} \left[\frac12 \left(S^+_{f\mathbf{R}} S^-_{f\mathbf{R}'} e^{i2\phi_f} + S^-_{f\mathbf{R}} S^+_{f\mathbf{R}'} e^{-i2\phi_f}\right) + \left(S^z_{f\mathbf{R}} S^z_{f\mathbf{R}'} - \frac14 n_{f\mathbf{R}} n_{f\mathbf{R}'}\right)\right]\,,
\label{eqn:HJ_real}
\end{align}
in which the anti-ferromagnetic correlation is only considered between the $f$ orbitals. This form is equivalent to Eqs.~(\ref{eq:kinetic}) and (\ref{eq:interaction}) in the main text. 

\section{Pairing channels, BdG Hamiltonian and Energetics of the Pairing States}
In this section, we provide a more detailed description of our analysis on the superconducting states of the two effective $t$-$J$ models discussed in the main text. 

We decouple the effective interaction in the pairing channels following  the approaches of Ref.~\cite{Goswami2010Superconductivity,Yu2013FeSC}. 
Accordingly, the superconducting order parameters can decompose into irreducible representations of the $C_3$ point group: the non-degenerate  $A_1$ channel and the two-fold degenerate $^{1,2}E$ channels. 
The resulting Bogoliubov-de Gennes (BdG) Hamiltonian takes the  form
\begin{equation}
H_{mf}=\sum_{\mathbf{k}}\Psi_{\mathbf{k}}^{\dagger}
\left[\begin{array}{cc}\mathbf{h}_{\mathbf{k}}
 & \mathbf{\Delta}_{\mathbf{k}}\\
\mathbf{\Delta}^{*}_{\mathbf{k}} &-\mathbf{h}_{\mathbf{k}}
\end{array}\right]\Psi_{\mathbf{k}}+\sum_{\mathbf{k},j}\mathcal{E}_{\mathbf{k}j}+\frac{3N}{2J}\sum_{\Gamma}|\Delta_\Gamma|^2\,,
\end{equation}
where the Nambu spinor is defined as:
$\Psi^{\dagger}_{\mathbf{k}}=(f^{\dagger}_{{\mathbf{k}}\uparrow}, c^{\dagger}_{{\mathbf{k}}\uparrow},f_{-{\mathbf{k}}\downarrow},c_{-{\mathbf{k}}\downarrow})$, and $h_\mathbf{k}$ is the normal-state Hamiltonian in the $(f,c)$ orbital basis \cite{xie2025kondolattice,Xie2024Superconductivity}. 
Since pairing arises only among the localized $f$-electrons, the pairing matrix $\mathbf{\Delta}_{\mathbf{k}}$ is nonzero solely in the $f$-orbital sector
 $\mathbf{\Delta}_{\mathbf{k}}
=\mathrm{diag}[\sum_\Gamma\Delta_\Gamma\gamma_\Gamma(\mathbf{k}),0]$, with $\Delta_\Gamma=-\frac{2J}{3N}\sum_k \langle\gamma_\Gamma(\mathbf{k})f_{-\mathbf{k}\downarrow}f_{\mathbf{k}\uparrow}\rangle$. 

Before discussing the superconducting properties, we first clarify the order parameter symmetry information to build intuition about the symmetry of the pairing states.
We define the spin-singlet and triplet pairing operators on the nearest-neighbor bonds of the more localized $f$-orbital, $\Delta_{j}^{s,t}=\langle f_{i\uparrow}f_{i+\bm{\delta}_j \downarrow}\mp f_{i\downarrow}f_{i+\bm{\delta}_j \uparrow}\rangle$, where $j=1,2,3$ labels the three nearest-neighbor bond directions $\bm{\delta}_j$, and the upper (lower) sign corresponds to the singlet (triplet) channel. 
The pairing amplitude $\Delta_\Gamma$ in each irreducible representation $\Gamma=A_1, {}^1E, {}^2E$ can then be expressed in terms of these bond pairing $\Delta_{j}^{s,t}$ as:
\begin{align}
    \Delta_\Gamma=&-\frac{J}{3}\cos{\phi_f}(\Delta_1^s+\omega_\Gamma\Delta_2^s+\omega_\Gamma^2\Delta_3^s)\nonumber\\
    &+\frac{J}{3}\sin{\phi_f}(\Delta_1^t+\omega_\Gamma\Delta_2^t+\omega_\Gamma^2\Delta_3^t)\,,
\end{align}
where $\omega_\Gamma=1, e^{i2\pi/3}, e^{i4\pi/3}$ corresponds to the basis functions of the $C_3$ irreducible representations.
Due to the presence of strong spin-orbit coupling (SOC), spin SU(2) symmetry is explicitly broken. 
As a result, the superconducting order parameter is no longer constrained to be an even or odd function of $\vk$, and spin-singlet and spin-triplet components are allowed to mix within each irreducible representations of the  $C_3$ symmetry group.
In particular, in the model with valley Chern number $\mathcal{C}=(-1,-1)$, when the displacement field vanishes ($\phi_f=\pi$ in this case), the order parameter becomes an even function of $\vk$, leading to pure ``spin-singlet'' pairing. 
This is consistent with prior results on one-band $t$-$J$ models on triangular lattices with real hoppings, where the dominant pairing channel is found to be of singlet $d_{x^2-y^2}+id_{xy}$ character. \cite{Baskaran2003Electronic,Kumar2003Superconductivity,Wang2004Doped,Ogata2003Superconducting, Watanabe2004Variational,Koretsune2005Pairing}. 

To determine which pairing channel is energetically favorable, we compute the ground state energy density as a function of pairing amplitudes
\begin{eqnarray}
E_g=\sum_{{\Gamma}}\frac{3N}{2J}|\Delta_{\Gamma}|^2-\sum_{{\mathbf{k}},j=\pm}(E_{{\mathbf{k}},j}-\mathcal{E}_{{\mathbf{k}},j}),
\label{eq:free_energy}
\end{eqnarray}
by minimizing it with respect to each pairing amplitude $\Delta_{{\Gamma}}$. 
Here $E_{{\mathbf{k}},j}$ denote the quasiparticle dispersion spectra obtained by diagonalizing the Hamiltonian $H_{mf}$, and $\mathcal{E}_{{\mathbf{k}},j}$ is the corresponding normal-state energy.
Note that the ground state energies contain the contributions from the combined $f$ and $c$ electrons.
In the following, we analyze the features of the unconventional superconducting state realized in the system. 
Throughout this work, we scale all energies by the bandwidth evaluated at $\varepsilon_D=10$ meV, and the chemical potential $\mu$ is tuned to maintain half-filling unless explicitly stated otherwise.
These conventions are applied consistently to both models. 

\begin{figure}[t]
    \centering
    \includegraphics[width =0.7\linewidth]{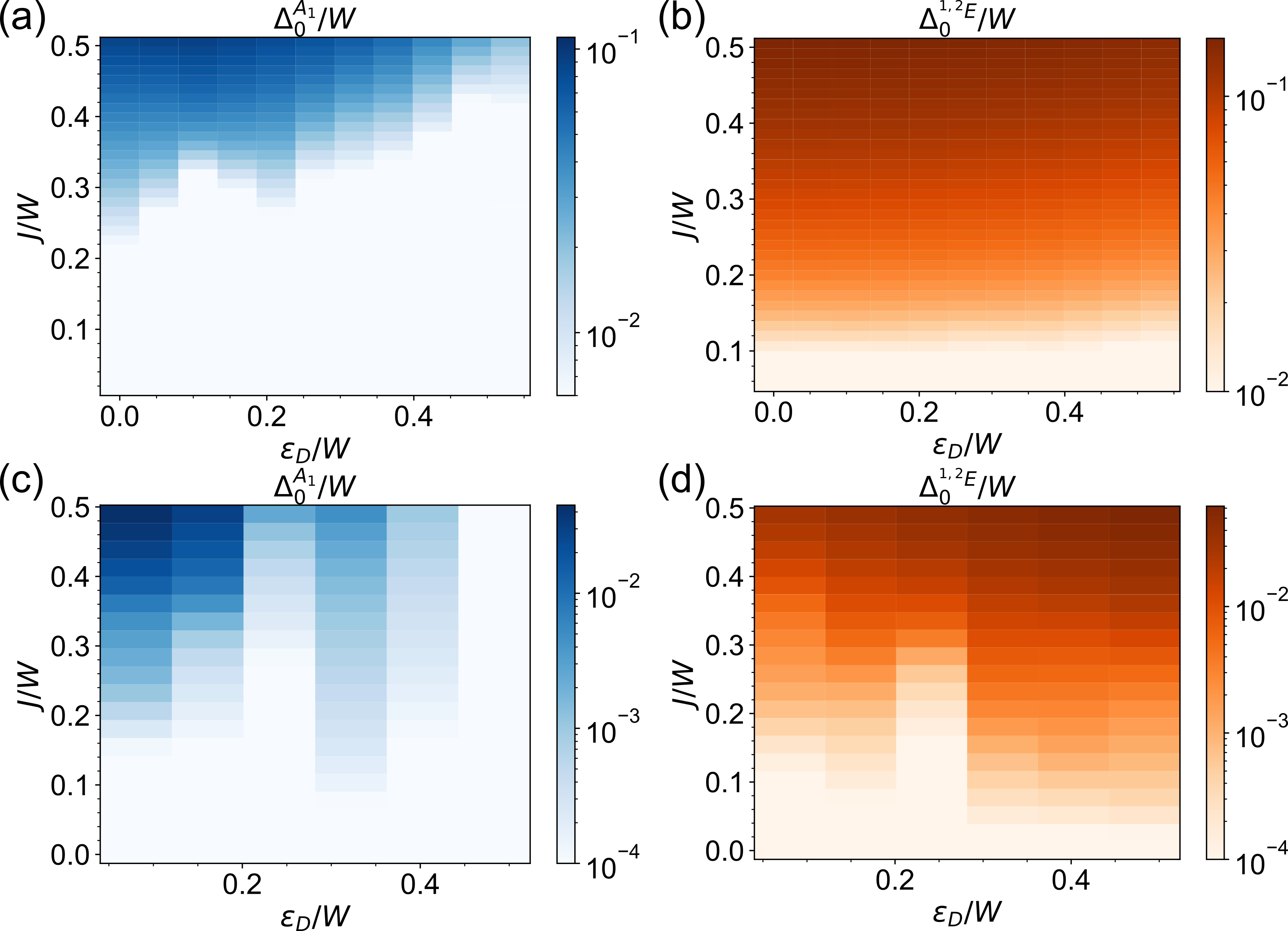}~
    \caption{Pairing amplitudes from the effective $t$-$J$ models. The color maps show the superconducting order parameter  $\Delta_0$ (in log scale) for the $A_1$ and ${}^{1,2}E$ pairing channels as a function of  the displacement field $\varepsilon_D$ and the exchange coupling $J$. 
    (a)(c) correspond to the model with the same valley Chern numbers $\mathcal{C}=(-1,-1)$, while (b)(d) are for $\mathcal{C}=(-1,+1)$. 
    For each plot, the pairing amplitudes of other channels are set to zero. }
    \label{fig:pairing_amp_A_E}
\end{figure}

The competition between different pairing symmetries is examined by comparing the pairing amplitudes of the  $A_1$ and ${}^{1,2}E$ irrep channels. 
Fig.~\ref{fig:pairing_amp_A_E} illustrates the evolution of superconducting order parameters under variations of  exchange coupling $J$ and displacement field $\varepsilon_D$, with all other pairing channels fixed at zero. 
The color intensity is shown on a logarithmic scale to highlight the growth of superconducting order in regions with exponentially small but finite pairing. 
In both models, the ${}^{1,2}E$ pairing channel develops a larger superconducting amplitude than the $A_1$ channel, demonstrating that the former is energetically more favorable. 

\begin{figure}[t]
    \centering
    \includegraphics[width =0.7\linewidth]{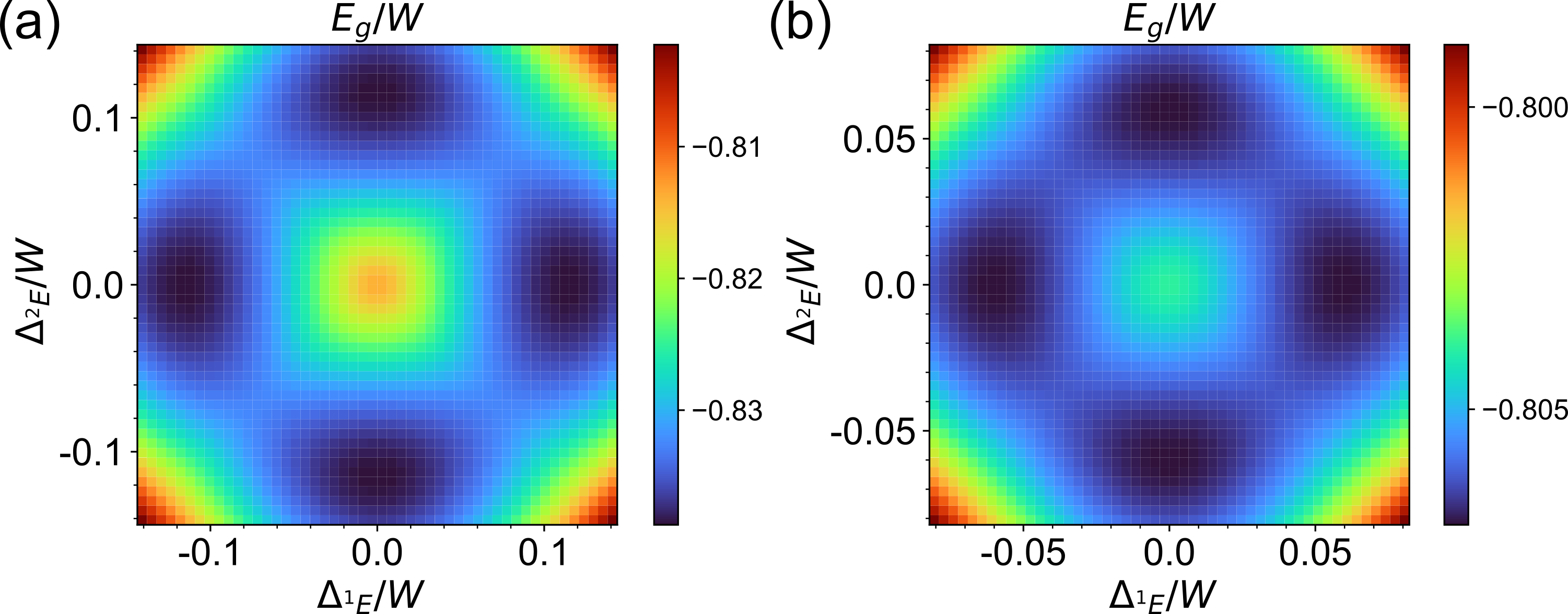}~
    \caption{Ground state energy density $E_g$ as a function of pairing amplitude $\Delta_{{}^1E}$ and $\Delta_{{}^2E}$ for the two models: (a) $\mathcal{C}=(-1,-1)$ (b) $\mathcal{C}=(-1,+1)$. The pairing amplitude of channel $A_1$ is set to be 0, and the relative phase between $\Delta_{{}^1E}$ and $\Delta_{{}^2E}$ is fixed at $\pi/2$.}
    \label{fig:Eg_1E2E_pi/2}
\end{figure}

We now turn to a fully relaxed minimization of Eq.~(\ref{eq:free_energy}), allowing all pairing amplitudes $\Delta_{A_1}$, $\Delta_{{}^1E}$ and $\Delta_{{}^2E}$, as well as their relative phases, to vary freely. 
The energy minimum is found to lie along either the pure ${}^1E$ or ${}^2E$ axis, with no evidence for energetically favorable mixing between the two components. 
To further illustrate the result, Fig.~\ref{fig:Eg_1E2E_pi/2} shows additional two-dimensional slices of $E_g$ for both models, computed at a fixed relative phase of $\pi/2$ between  $\Delta_{{}^1E}$ and $\Delta_{{}^2E}$ with $\Delta_{A_1}=0$. 
The energy landscapes remain qualitatively similar to the zero-phase case in Fig.~\ref{fig:Eg_1E2E}, confirming that the preference for a pure ${}^1E$ or ${}^2E$ state persists when complex linear combinations of the two components are allowed.

\section{Determination of topological index}\label{sec:topo-idx}

In this section, we will discuss how we can identify the topological index of the Bogoliubov bands in both models, especially for the two-band model with the topological power-law orbital.

We start from the Kondo-lattice model with the MLWF and TPLO as the active degrees of freedom, i.e., the case with $\mathcal{C} = (-1, -1)$.
Since there is no spin-parallel pairing terms in the $t$-$J$ model, the ``creation operator'' of a Bogoliubov quasiparticle has the following form:
\begin{equation}\label{eqn:bdg-operator}
    \hat{\beta}^\dagger_{\vk, n, \uparrow} = u_{f, n}(\vk) f^\dagger_{\vk, \uparrow} + u_{c, n}(\vk) c^\dagger_{\vk,\uparrow} + v_{f, n}(\vk) f_{-\vk, \downarrow} + v_{c, n}(\vk) c_{-\vk,\downarrow}\,,
\end{equation}
in which $c^\dagger$ and $f^\dagger$ stand for the fermion creation operator for TPLO and MLWF, respectively.
We note that due to the $S_z$ conservation, $\hat{\beta}^\dagger$ operators will only contain the creation operators of the spin-up states, and annihilation operators of the spin-down states.
The Bogoliubov amplitudes $u, v$ satisfy the normalization condition:
\begin{equation}
    1 = \sum_{\alpha=c,f} u^*_{\alpha, n}(\vk) u_{\alpha,n}(\vk) + \sum_{\alpha=c,f} v^*_{\alpha,n}(\vk) v_{\alpha, n}(\vk)\,.
\end{equation}
Since the TPLO cannot be considered as a localized orbital, the natural choice for defining the Berry connection of the BdG states is the plane wave basis.
In this basis, the quasiparticle operator in Eq.~(\ref{eqn:bdg-operator}) can be rewritten as:
\begin{align}
    \hat{\beta}^\dagger_{\vk, n, \uparrow} =&
    \sum_{\mathbf{Q}}\left[c^\dagger_{\vk - \mathbf{Q},\uparrow}\left(u_{f, n}(\vk) \tilde{u}^{(\uparrow)}_{\mathbf{Q}, f}(\vk) + u_{c, n}(\vk) \tilde{u}^{(\uparrow)}_{\mathbf{Q},c}(\vk) \right) + c_{-\vk - \mathbf{Q},\downarrow}\left(v_{f, n}(\vk) \tilde{u}^{(\downarrow)*}_{\mathbf{Q},f}(-\vk) + v_{c, n}(\vk) \tilde{u}^{(\downarrow)*}_{\mathbf{Q},c}(-\vk)\right)\right]\,,
\end{align}
in which $c^\dagger_{\mathbf{k} - \mathbf{Q},\sigma}$ creates a plane wave with wave vector $\vk - \mathbf{Q}$ in the spin-valley sector $\sigma$.
As such, an off-diagonal (non-Abelian) Berry connection of BdG states can be defined in the following form:
\begin{align}
    &\left[\exp\left(i\mathbf{A}(\vk) \cdot d\vk \right)\right]_{mn}\nonumber \\
    =& \sum_{\mathbf{Q}} \left(u^*_{f, n}(\vk) \tilde{u}^{(\uparrow)*}_{\mathbf{Q}, f}(\vk) + u^*_{c, n}(\vk) \tilde{u}^{(\uparrow)*}_{\mathbf{Q},c}(\vk) \right)\left(u_{f, n}(\vk + d\vk) \tilde{u}^{(\uparrow)}_{\mathbf{Q}, f}(\vk + d\vk) + u_{c, n}(\vk + d\vk) \tilde{u}^{(\uparrow)}_{\mathbf{Q},c}(\vk + d\vk) \right) \nonumber \\
    & + \left(v^*_{f, n}(\vk) \tilde{u}^{(\downarrow)}_{\mathbf{Q},f}(-\vk) + v^*_{c, n}(\vk) \tilde{u}^{(\downarrow)}_{\mathbf{Q},c}(-\vk)\right)\left(v_{f, n}(\vk + d\vk) \tilde{u}^{(\downarrow)*}_{\mathbf{Q},f}(-\vk - d\vk) + v_{c, n}(\vk + d\vk) \tilde{u}^{(\downarrow)*}_{\mathbf{Q},c}(-\vk - d\vk)\right)\,.
\end{align}
Accordingly, a Wilson loop can be defined as the path order integral of the non-Abelian connection along a given direction.
For example, we can choose the Wilson loop contour along $\mathbf{b}_2$ with fixed value of $k_1$:
\begin{equation}\label{eqn:bdg-wilson}
    W(k_1) = \mathcal{P}\exp\left(i\int_0^{2\pi} dk_2 \, A_2(\vk)\right) \approx \prod_{K_2=0}^{L_2-1} \exp\left[ i \mathbf{A}\left(\frac{k_1}{2\pi}\mathbf{b}_1 + \frac{K_2}{L_2} \mathbf{b_2}\right) \cdot \frac{\mathbf{b}_2 }{L} \right]\,,
\end{equation}
which is a unitary matrix.
Its eigenvalues $\lambda_j(k_1) = e^{i\theta_j(k_1)}$ and its determinant are gauge invariant quantities, which carry important information about the topology of the Bogoliubov quasiparticles.

\begin{figure}
    \centering
    \includegraphics[width=\linewidth]{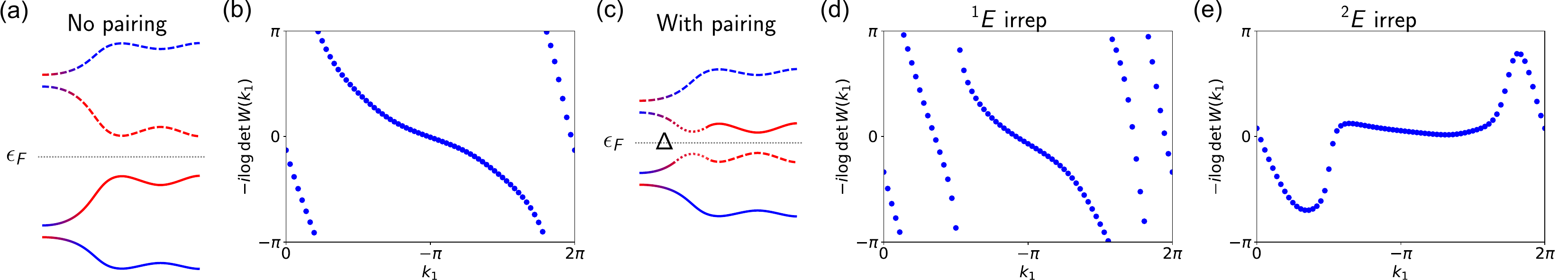}
    \caption{(a) Schematic of the Bogoliubov-de-Gennes quasiparticle spectrum without superconductivity pairing terms in the parameter regime with the same valley Chern number in the top two moir\'e bands.
    Here we use solid and dashed lines to represent the particle and hole sectors, respectively.
    (b) The Wilson loop of the two positive BdG bands in the case (a). Here the winding number $\mathcal{C} = -2$ is inherited from the non-interacting band structure of this system.
    (c) Schematic of the BdG quasiparticle spectrum with Cooper pairing in the top most moir\'e band.
    The pairing term introduces a ``hybridization'' between the particle and the hole sectors.
    (d-e) The Wilson loop of the two positive BdG bands in the presence of a pairing order parameter with $C_{3}$ irrep $^1E$ and $^2E$, respectively.
    The total Chern number change by $\Delta \mathcal{C} = \pm 2$.
    Here we used the model parameters obtained at $\varepsilon_D = 10\rm \,meV$.
    }
    \label{fig:partial-wannier-wilson-loop}
\end{figure}

Since the the TPLO itself is topological obstructed, the determination of the topology of the BdG state might be affect. 
Therefore, a reasonable ``reference point'', which corresponds to the case in the absence of Cooper pairing, has to be considered.
Here, we consider the case with both bands fully filled and no pairing terms as the reference state.
This state is sketched in Fig.~\ref{fig:partial-wannier-wilson-loop}(a), where the solid and dashed lines stand for the particle and hole sectors of the BdG bands.
The phase of $\det W(k_1)$ computed from the top two bands is shown in Fig.~\ref{fig:partial-wannier-wilson-loop}(b).
The non-vanishing winding number $\mathcal{C} = -2$ is clearly inherited from the non-interacting band structure.
When the electrons form Cooper pairs, the particle and hole sectors will be ``hybridized'', as sketched in Fig.~\ref{fig:partial-wannier-wilson-loop}(c).
When the pairing function form ${}^1 E$ or ${}^2 E$ representations of $C_3$, the corresponding plots of $\det W(k_1)$ are shown in Figs.~\ref{fig:partial-wannier-wilson-loop}(d-e), respectively.
Here, we have used the order parameter amplitude $\Delta/W \approx 0.15$.
Compared to their counterparts without Cooper pairing, the Wilson loop winding number changes by $\Delta \mathcal{C} = \pm 2$ in these two cases, which is indeed a hallmark of $d\pm id$ pairing symmetry.

The subtraction of winding numbers of ``normal state'' can be also considered from another perspective. 
When we truncate the fully-filled remote bands of the continuum model during the projection procedure, these bands will carry the opposite winding number as the two active bands.
In the presence of pairing order parameter, these bands will still remain filled.
Subtracting the winding numbers of the normal state from the superconducting state is equivalent to adding the total winding number of all the filled bands.
Therefore, the Chern number change $\Delta \mathcal{C} = \pm 2$ is indeed the topological index of the superconducting state with all bands considered.

\begin{figure}
    \centering
    \includegraphics[width=0.75\linewidth]{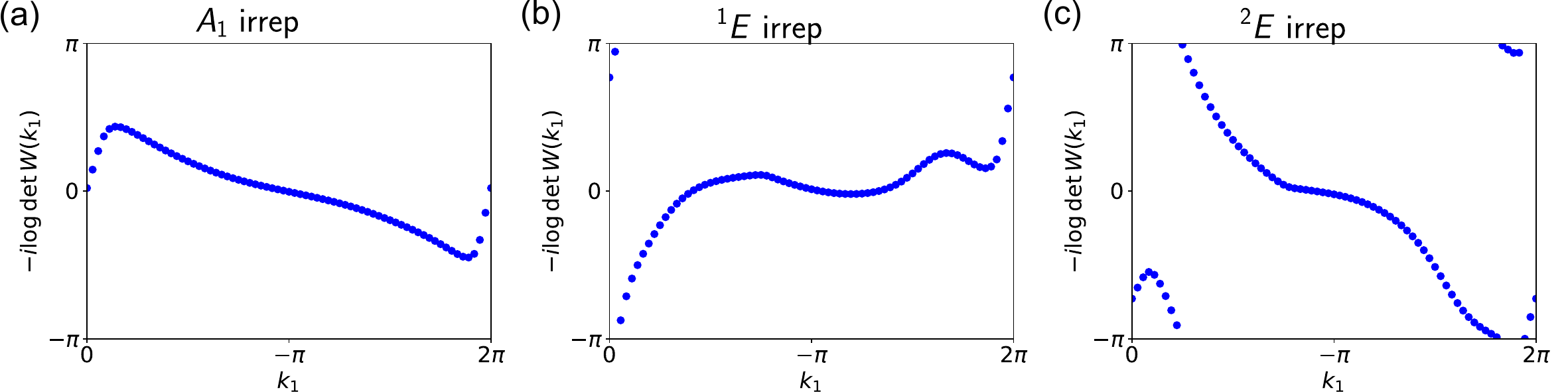}
    \caption{The phase of the determinant of the Bogoliubov Wilson loop in the parameter regime with $\mathcal{C} = (-1, + 1)$ and pairing order parameter (a) $A_1$ irrep, (b) $^1E$ irrep and (c) $^2E$ irrep.
    Here parameters are obtained from the model with displacement field potential $\varepsilon_D = 15\rm\, meV$.
    }
    \label{fig:winding-two-orb}
\end{figure}

The generalization to the two orbital model with Chern numbers $\mathcal{C} = (-1, +1)$ is straightforward.
In such case, the Wilson loops can be computed similarly, by simply replacing the plane wave decomposition of MLWF and the TPLO with their counterparts for the two exponentially localized orbitals.
Fig.~\ref{fig:winding-two-orb} shows the winding of $\det W(k_1)$ with three types of order parameters with different $C_3$ irreps.
In particular, $^{1,2}E$ irreps can lead to topological superconductors with Chern number $\mathcal{C} = \pm 1$.

\begin{figure}
    \centering
    \includegraphics[width=0.75\linewidth]{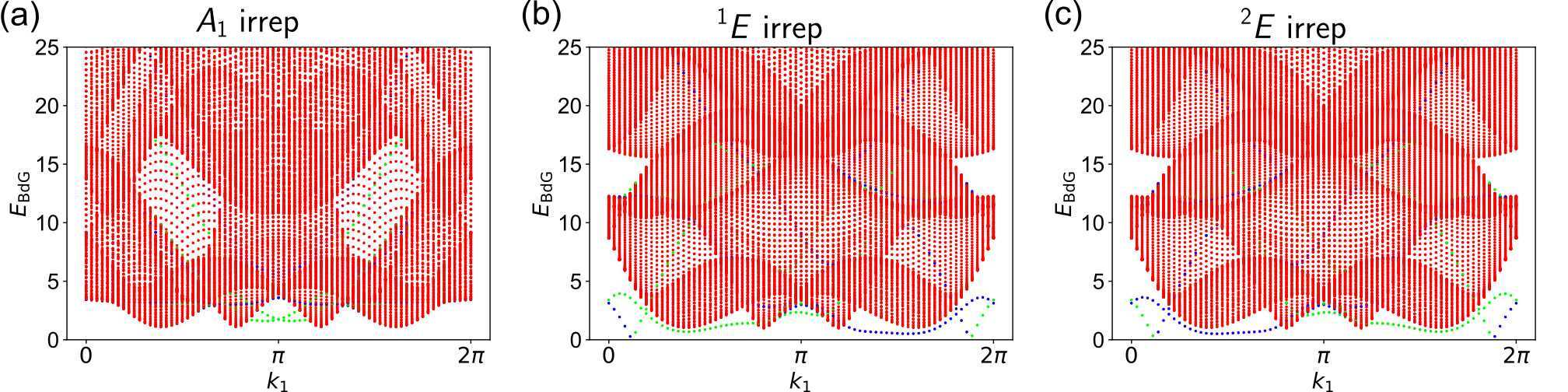}
    \caption{The edge modes of the Bogoliubov quasiparticle spectrum with different types of pairing order parameters in the parameter regime with $\mathcal{C} = (-1, +1)$.
    Here we use red, blue and green markers to represent states located in the bulk, along the top edge and bottom edge of the slab, respectively.
    }
    \label{fig:edge}
\end{figure}

Since the effective model in the parameter regime with $\mathcal{C} = (-1, +1)$ is a legitimate tight-binding model, we are also able to perform an open boundary condition calculation for these Bogoliubov quasiparticles.
For three types of pairing order parameters, the open boundary condition spectra are shown Fig.~\ref{fig:edge}.
Here we computed the open boundary Bogoliubov spectra using particle-hole redundant representation, and the number of edge modes on each edge is twice the Chern number suggested by the Wilson loop winding number.

\end{document}